\documentclass[aps,prx,twocolumn,showpacs]{revtex4-1}
\usepackage{bm}
\usepackage{mathrsfs}
\usepackage{amsmath}
\usepackage{amssymb}
\usepackage{graphicx}
\usepackage{amsfonts}
\usepackage{amsthm}
\usepackage{color}
\usepackage{dcolumn}
\usepackage{txfonts}
\usepackage[caption=false]{subfig}
\DeclareMathOperator{\Cov}{Cov}

\begin{document}

\title{Three-Axis Spin Squeezed States Associated with Excited-State Quantum Phase Transitions}
\author{Chon-Fai Kam}
\email{Email: dubussygauss@gmail.com}
\affiliation{Quantum Theory Group, Dipartimento di Fisica e Chimica Emilio Segrè,\\
Università degli Studi di Palermo, Via Archirafi 36, I-90123 Palermo, Italy}

\begin{abstract}
Spin squeezing in collective atomic ensembles enables quantum-enhanced metrology by suppressing noise below the standard quantum limit via nonlinear interactions; extending Kitagawa and Ueda's one- and two-axis twisting paradigms, we introduce three-axis spin squeezed states within the anisotropic Lipkin-Meshkov-Glick (LMG) model, featuring direction-dependent quadratic couplings and external fields that unify uniaxial and biaxial regimes into asymmetric quantum rotors with elliptical quasiprobability distributions and multipartite entanglement. Deriving the Hamiltonian and semiclassical Euler-top dynamics, we uncover optimal squeezing scalings of $\xi^2 \sim N^{-2/3}$ for one-axis twisting and Heisenberg-limited $\xi^2 \sim 1/N$ for two-axis variants, while three-axis states—analyzed via Majorana representations and Husimi-$Q$ functions—boost metrological gain and concurrence up to $\sin|\xi| \approx 1$ for low-$j$ systems. Crucially, anisotropy tuning drives second-order ground-state quantum phase transitions between paramagnetic and ferromagnetic phases, with critical exponents $\gamma=1$ and Kibble-Zurek universality in quench dynamics at $\lambda_z = \lambda_{x,y}$, advancing many-body theory and proposing implementations in Rydberg arrays and cavity-QED for precision sensing and critical simulation.
\end{abstract}

\maketitle

\section{Introduction}
Since the seminal work by Kitagawa and Ueda on squeezed spin states~\cite{kitagawa1993squeezed}, spin squeezing in collective atomic ensembles has emerged as a foundational concept in quantum optics and many-body physics. This phenomenon enables quantum-enhanced metrology by suppressing quantum noise below the standard quantum limit through nonlinear interactions, with applications in precision sensing, quantum information processing, and fundamental tests of quantum mechanics. Analogous to quadrature squeezing in bosonic systems---where the variance in one quadrature is reduced below the vacuum level of $1/4$ while maintaining the Heisenberg uncertainty relation---spin squeezing manifests in ensembles of $N=2j$ spin-$1/2$ particles, equivalent to a collective spin-$j$ system. A spin coherent state $|\theta, \phi\rangle$ corresponds to $N$ identical spins aligned along the direction $(\theta, \phi)$ on the Bloch sphere, exhibiting isotropic quasiprobability distributions and Poissonian fluctuations scaling as $\sqrt{N}$.
In contrast, spin squeezed states feature elliptical quasiprobability distributions, reflecting anisotropic noise reduction driven by inter-particle correlations and multipartite entanglement. Linear Hamiltonians, which merely rotate the collective spin without generating correlations, are insufficient for squeezing; nonlinear interactions are essential. Kitagawa and Ueda identified two paradigmatic mechanisms: one-axis twisting (OAT), which shears the uncertainty region via a uniaxial quadratic interaction, and two-axis twisting (TAT), which achieves more efficient squeezing through biaxial couplings.

This framework extends beyond noise suppression, revealing profound connections to quantum phase transitions (QPTs) and classical-quantum correspondences. The Lipkin-Meshkov-Glick (LMG) model~\cite{lipkin1965validity,meshkov1965validity,glick1965validity}, originally developed in nuclear physics to describe infinite-range spin interactions, provides an ideal platform for exploring these links. In its generalized anisotropic form, the LMG Hamiltonian is
$$H = -\frac{1}{N} \sum_{\alpha=x,y,z} \lambda_\alpha (S_\alpha)^2 - \sum_{\alpha=x,y,z} h_\alpha S_\alpha,$$
where $S_\alpha = \sum_{i=1}^N \sigma_\alpha^{(i)}/2$ are collective spin operators, $\lambda_\alpha$ denote direction-dependent coupling strengths, and $h_\alpha$ are external magnetic fields. The $1/N$ prefactor ensures a well-defined thermodynamic limit as $N \to \infty$. Anisotropy in $\lambda_\alpha$ enables diverse regimes; for example, $\lambda_z = 0$ and $\lambda_x = \lambda_y$ restricts interactions to the $xy$-plane, while higher-order terms like $(S_\alpha)^4$ can enrich the dynamics. Here, we focus on the quadratic form central to spin squeezing and QPTs.

In the semiclassical large-$N$ limit, the collective spins behave as a classical vector $\mathbf{S} = (S_x, S_y, S_z)$ with Poisson brackets ${S_\alpha, S_\beta} = \epsilon_{\alpha\beta\gamma} S_\gamma$, leading to equations of motion
$$\frac{dS_\alpha}{dt} = \epsilon_{\alpha\beta\gamma} S_\beta \left( \frac{2\lambda_\gamma}{N} S_\gamma + h_\gamma \right),$$
analogous to the Euler equations for an asymmetric rigid body (Euler top), with $\lambda_\alpha$ inversely related to moments of inertia.
The quadratic interactions generate spin squeezing, quantified by Kitagawa and Ueda's parameter $\xi^2 = (N/\langle \mathbf{S} \rangle^2) \min_{\mathbf{n}\perp} \langle (\Delta S{\mathbf{n}\perp})^2 \rangle$, where $\xi^2 < 1$ indicates squeezing below the SQL, and $\mathbf{n}\perp$ is perpendicular to the mean spin direction. For OAT ($\lambda_z \neq 0$, $\lambda_x = \lambda_y = 0$, $h_\alpha = 0$), the Hamiltonian is $H_{\rm OAT} = -\lambda_z S_z^2 / N$, yielding optimal squeezing $\xi_{\min}^2 \propto N^{-2/3}$ at evolution time $t \propto N^{-1/3}$. For two-axis counter-twisting (TACT), $H_{\rm TACT} = \chi (S_x^2 - S_y^2)$ (or equivalent forms with $S_x S_y + S_y S_x$), the scaling approaches the Heisenberg limit, $\xi_{\min}^2 \propto N^{-1}$. Semiclassically, OAT deforms phase-space trajectories uniaxially, while TACT induces biaxial distortions akin to an asymmetric top.

Extending these paradigms, we introduce three-axis spin squeezed states within the anisotropic LMG model, unifying OAT and TAT regimes into asymmetric quantum rotors with tunable ellipticities. By deriving the Hamiltonian, semiclassical dynamics, and quasiprobability representations (Majorana stars and Husimi-$Q$ functions), we reveal enhanced metrological gains, with $\xi^2 \sim N^{-2/3}$ for OAT-like cases and $\xi^2 \sim N^{-1}$ for TAT variants, alongside boosted concurrence up to $\sin|\xi| \approx 1$ in low-$j$ systems.
Beyond squeezing, the model exhibits rich critical phenomena. Tuning $\lambda_\alpha$ and $h_\alpha$ induces second-order ground-state QPTs from paramagnetic to ferromagnetic phases, with critical exponents $\gamma=1$ and Kibble-Zurek scaling in quench dynamics at $\lambda_z = \lambda_{x,y}$. Excited-state QPTs manifest as singularities in the density of states or level clustering. The model is integrable in limits like isotropic couplings (solvable via Bethe ansatz) or vanishing $\lambda_\alpha$, but extensions---such as breaking all-to-all symmetry or adding degrees of freedom---introduce non-integrability and potential quantum chaos, probed via level statistics or out-of-time-order correlators.

In summary, the generalized LMG model bridges spin squeezing, rotational analogies, and critical phenomena, advancing many-body theory. We propose implementations in Rydberg atom arrays and cavity-QED platforms for precision sensing and quantum simulation of criticality.

\section{The Model}
We consider a generalized quadratic spin Hamiltonian that extends the one- and two-axis twisting models to incorporate three-axis interactions within the anisotropic Lipkin-Meshkov-Glick (LMG) framework. The Hamiltonian is given by
\begin{equation}\label{SpinHamiltonian}
H = \frac{\chi_0}{2} (\mathbf{J}^2 - J_z^2) + \frac{\chi_1}{2} (J_x^2 - J_y^2) + \frac{\chi_2}{2} (J_x J_y + J_y J_x),
\end{equation}
where $\mathbf{J}^2 \equiv J_x^2 + J_y^2 + J_z^2$, and the collective spin operators $J_x$, $J_y$, $J_z$ satisfy the standard angular momentum commutation relations $[J_i, J_j] = i \epsilon_{ijk} J_k$ (with $\hbar = 1$). Here, $\chi_0$, $\chi_1$, and $\chi_2$ are tunable coupling strengths with dimensions of frequency. Note that the corresponding classical Hamiltonian, including its classical bifurcation structure at the phase transition and its realization in nonlinear optics, has been studied previously \cite{kam2025analogues, kam2025realization}

The first term, $\frac{\chi_0}{2} (\mathbf{J}^2 - J_z^2) = -\frac{\chi_0}{2} J_z^2 + \frac{\chi_0}{2} j(j+1)$ (up to a constant shift), corresponds to one-axis twisting (OAT) along the $z$-axis, which shears the quasiprobability distribution of an initial spin coherent state~\cite{kitagawa1993squeezed}. The second term, $\frac{\chi_1}{2} (J_x^2 - J_y^2) = \frac{\chi_1}{4} (J_+^2 + J_-^2)$, and the third term, $\frac{\chi_2}{2} (J_x J_y + J_y J_x) = \frac{\chi_2}{4i} (J_+^2 - J_-^2)$, represent two-axis twisting (TAT) mechanisms. These induce simultaneous clockwise and counterclockwise twisting about orthonormal axes in the $xy$-plane (for $\chi_1$) or rotated by $\pm \pi/4$ (for $\chi_2$), where $J_\pm \equiv J_x \pm i J_y$ are the ladder operators.
To diagonalize the TAT contributions and reveal the underlying structure, we perform a rotation about the $z$-axis using the operator $R_z(\theta) = e^{-i \theta J_z}$, yielding transformed operators
\begin{subequations}
\begin{align}
I_x &\equiv R_z^\dagger(\theta) J_x R_z(\theta) = \cos\theta , J_x - \sin\theta , J_y, \\
I_y &\equiv R_z^\dagger(\theta) J_y R_z(\theta) = \sin\theta , J_x + \cos\theta , J_y, \\
I_z &\equiv R_z^\dagger(\theta) J_z R_z(\theta) = J_z.
\end{align}
\end{subequations}
The rotated Hamiltonian $H' \equiv R_z^\dagger(\theta) H R_z(\theta)$ becomes
\begin{equation}\label{RotatedHamiltonian}
H' = \frac{\chi_0}{2} (\mathbf{I}^2 - I_z^2) + \frac{\chi}{2} (I_x^2 - I_y^2),
\end{equation}
where $\theta = \arg(\chi_1 - i \chi_2)/2$ and $\chi = \sqrt{\chi_1^2 + \chi_2^2}$. Assuming $\chi_0 > 0$ without loss of generality, the regime $\chi_0 > \chi$ is dominated by OAT, while $\chi > \chi_0$ emphasizes TAT. Equivalently,
\begin{equation}\label{QRotor}
H' = \frac{\chi_0 + \chi}{2} I_x^2 + \frac{\chi_0 - \chi}{2} I_y^2,
\end{equation}
interpreting the system as an asymmetric quantum rotor with moments of inertia $1/(\chi_0 + \chi)$ and $1/(\chi_0 - \chi)$ about the $x$- and $y$-axes, respectively, when $\chi_0 > \chi$~\cite{vidal2004finite}.

The associated three-axis spin squeezed states, evolving from an initial spin coherent state $|j, \mathbf{n}_0\rangle$ under $H$, are defined as
\begin{equation}
|j, \mathbf{n}_0, \boldsymbol{\mu}\rangle \equiv \exp\left\{ -\frac{i}{2} \left[ \mu_0 (\mathbf{J}^2 - J_z^2) + \frac{\xi}{2} J_+^2 + \frac{\xi^*}{2} J_-^2 \right] \right\} |j, \mathbf{n}_0\rangle,
\end{equation}
where $\boldsymbol{\mu} = (\mu_0, \mu_1, \mu_2) = (\chi_0 t, \chi_1 t, \chi_2 t)$, $\xi = \mu_1 - i \mu_2$, and the initial state is
\begin{equation}
|j, \mathbf{n}_0\rangle = (1 + |\tau|^2)^{-j} \sum_{m=-j}^j \sqrt{\binom{2j}{j+m}} \tau^{j+m} |j, m\rangle,
\end{equation}
with $\tau = \tan(\theta_0/2) e^{-i \phi_0}$ the stereographic projection of $\mathbf{n}_0 = (\theta_0, \phi_0)$ from the north pole~\cite{perelomov2012generalized}.
Since $J_\pm^2 |j, m\rangle \propto |j, m \pm 2\rangle$, the TAT terms preserve parity, implying that states evolved from even-parity initial conditions (e.g., lowest-weight $|j, -j\rangle$) span only even $m$-shifts. Thus, $|j, \boldsymbol{\mu}\rangle$ and $J_\pm^2 |j, \boldsymbol{\mu}\rangle$ are even-parity eigenstates of $(-1)^{J_z + j}$ with eigenvalue $+1$, while $J_\pm |j, \boldsymbol{\mu}\rangle$ are odd-parity with $-1$.

This model unifies OAT and TAT into a tunable three-axis framework, enabling asymmetric squeezing and connections to quantum rotors, as explored in subsequent sections through spectral analysis, phase-space representations, and entanglement measures.

\section{The spectrum of the tri-axis spin squeezed states}

In this section, we examine the energy spectrum of the tri-axis twisting Hamiltonian given in Eq.~\eqref{SpinHamiltonian}, focusing on its implications for the dynamics of tri-axis spin squeezed states and the emergence of excited-state quantum phase transitions (ESQPTs). The spectrum provides critical insights into the quantum many-body behavior of the system, revealing how tunable anisotropies in the coupling strengths $\chi_0$, $\chi_1$, and $\chi_2$ drive transitions between different squeezing regimes and critical phenomena in excited states.
For a collective spin system with total angular momentum $j$, the Hamiltonian acts on the $(2j+1)$-dimensional Hilbert space spanned by the Dicke states $|j, m\rangle$, where $m = -j, -j+1, \dots, j$. Due to the quadratic nature of the interactions and the parity conservation under two-axis twisting terms (as $J_\pm^2$ shifts $m$ by $\pm 2$), the matrix representation of $H$ decouples into even- and odd-parity subspaces. This block-diagonal structure facilitates exact diagonalization for small to moderate $j$, while for larger $j$, semiclassical approximations or numerical methods become essential.

To illustrate the spectral features, we consider the rotated Hamiltonian $H'$ in Eq.~\eqref{RotatedHamiltonian}, which simplifies analysis by diagonalizing the two-axis components. The eigenvalues $E_k$ ($k = 1, 2, \dots, 2j+1$) are obtained by solving the time-independent Schrödinger equation $H' |\psi_k\rangle = E_k |\psi_k\rangle$. In the large-$j$ limit, the spectrum can be approximated semiclassically using the effective potential derived from the quantum rotor interpretation in Eq.~\eqref{QRotor}, where the energy levels correspond to quantized rotations with moments of inertia determined by $\chi_0 \pm \chi$.

\begin{figure}[tbp]
\includegraphics[width=0.8\columnwidth]{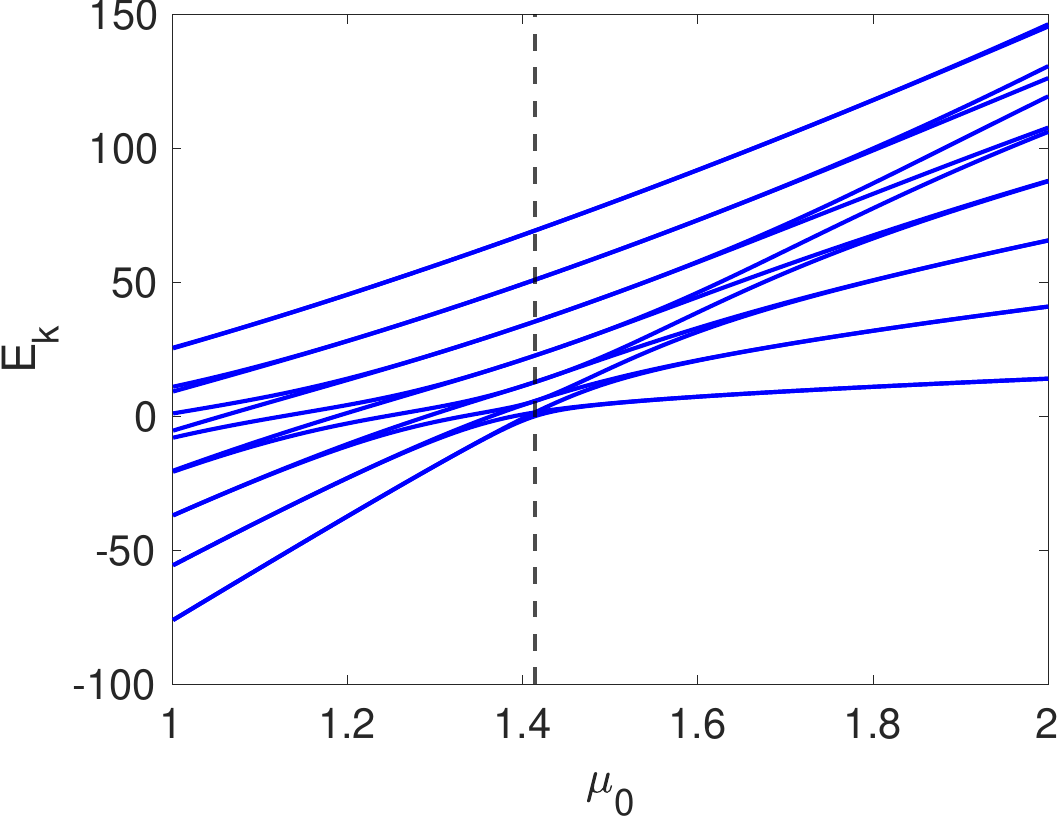}
\caption{Energy spectrum $E_k$ as a function of the normalized one-axis twisting parameter $\mu_0$ for $j=10$, illustrating an excited-state quantum phase transition (ESQPT) at $\mu_0^c \approx 1.5$ (vertical dashed line). The level clustering and avoided crossings highlight the transition from stable rotational modes to critical dynamics.}
\label{S1}
\end{figure}

Fig.~\ref{S1} displays the energy eigenvalues $E_k$ as a function of the normalized one-axis twisting parameter $\mu_0 = \chi_0 / \chi$ (assuming $\chi > 0$), computed numerically for a system with $j = 10$ (yielding 21 levels). The plot spans $\mu_0$ from 1.0 to 2.0, capturing the transition from two-axis-dominated ($\mu_0 < \sqrt{2}$) to one-axis-dominated ($\mu_0 > \sqrt{2}$) regimes. For $\mu_0 \lesssim 1.4$, the levels exhibit nearly linear dispersion with positive slopes for higher excited states and negative for lower ones, indicative of stable rotational modes in the asymmetric rotor picture. As $\mu_0$ increases, the levels begin to cluster and exhibit avoided crossings, particularly around the critical point $\mu_0^c \approx 1.5$ (marked by the vertical dashed line), where $\chi_0 = \chi$ and the effective moment of inertia about one axis diverges (cf. Eq.~\eqref{QRotor}).

This level clustering is a hallmark of an ESQPT, a quantum phase transition occurring in the excited-state manifold rather than the ground state~\cite{caprio2008excited, cejnar2021excited}. In the LMG model and its variants, ESQPTs manifest as singularities in the density of states $\rho(E) = \sum_k \delta(E - E_k)$, often scaling as $\rho(E) \sim |E - E_c|^{-\alpha}$ near the critical energy $E_c$, with exponent $\alpha$ depending on the system's dimensionality and symmetry~\cite{santos2016excited}. For the tri-axis case, the critical point at $\mu_0^c$ corresponds to a second-order ESQPT, where the semiclassical phase space undergoes a bifurcation from elliptic to hyperbolic fixed points~\cite{santos2016excited}, leading to a logarithmic divergence in $\rho(E)$ ($\alpha = 0$) in the thermodynamic limit $j \to \infty$.

Mathematically, near the ESQPT, the level spacing statistics transition from Poissonian (integrable regime) to Wigner-Dyson (chaotic regime), quantifiable via the nearest-neighbor spacing distribution $P(s)$. For small spacings, $P(s) \propto s^\beta$, with $\beta = 0$ (Poisson) away from criticality and $\beta = 1$ (orthogonal ensemble) at the transition. In our spectrum, the bunching of levels around $E \approx 0$ for $\mu_0 > \mu_0^c$ signals enhanced quantum fluctuations, which amplify squeezing in tri-axis states by facilitating multipartite entanglement across multiple energy scales.

The ESQPT also influences the time evolution of spin squeezed states. Starting from a coherent state $|j, \mathbf{n}_0\rangle$, the wavepacket spreads under $H$, with dephasing accelerated near criticality due to the dense spectrum. This leads to dynamical signatures such as power-law decay in survival probabilities $P(t) = |\langle \psi(0) | \psi(t) \rangle|^2 \sim t^{-\gamma}$, with $\gamma \approx 1$ at the ESQPT~\cite{relano2008decoherence, santos2016excited}. For metrological applications, operating near this critical point enhances sensitivity, as the squeezing parameter $\xi^2$ approaches the Heisenberg limit more rapidly, albeit at the cost of increased decoherence sensitivity.

In summary, the spectrum of the tri-axis Hamiltonian reveals a rich interplay between twisting mechanisms and critical dynamics, with the ESQPT at $\mu_0^c$ serving as a tunable knob for optimizing quantum-enhanced protocols~\cite{santos2016excited}. Future studies could explore finite-size scaling and chaotic indicators to further characterize this transition.

\section{The Majorana Star Representation}
The Majorana stellar representation provides a powerful geometric visualization of pure quantum states in finite-dimensional Hilbert spaces, particularly useful for symmetric multiqubit systems and spin-$j$ states~\cite{majorana1932atomi, bloch1946nuclear}. By mapping the state to a constellation of $2j$ points (Majorana stars) on the unit sphere, this approach reveals symmetries, entanglement structures, and phase-space distributions that are otherwise obscured in the standard basis \cite{kam2021berry, kam2020three}. For spin squeezed states, which exhibit anisotropic noise and multipartite correlations, the Majorana representation highlights the deformation of the initial coherent state's isotropic distribution into clustered or spread-out star configurations, offering insights into squeezing mechanisms and metrological utility~\cite{wick2004majorana, chabaud2017stellar}.

For a general spin-$j$ state $|\psi_j\rangle = \sum_{m=-j}^j c_m |j, m\rangle$, the overlap with a spin coherent state $|j, \mathbf{n}\rangle$ is
\begin{equation}
\langle j, \mathbf{n} | \psi_j \rangle = \sum_{m=-j}^j \sqrt{\binom{2j}{j+m}} \left( \cos\frac{\theta}{2} \right)^{j-m} \left( \sin\frac{\theta}{2} e^{i\phi} \right)^{j+m} c_m,
\end{equation}
where $\mathbf{n} = (\theta, \phi)$ parameterizes the direction on the Bloch sphere. The overlap with the antipodal coherent state $|j, -\mathbf{n}\rangle$ yields
\begin{subequations}
\begin{align}
\langle j, -\mathbf{n} | \psi_j \rangle &= \left( \sin\frac{\theta}{2} \right)^{2j} P_N(z), \\
P_N(z) &= \sum_{m=-j}^j (-1)^{j+m} \sqrt{\binom{2j}{j+m}} c_m z^{j+m},
\end{align}
\end{subequations}
with $N = 2j$ and $z = \cot(\theta/2) e^{i\phi}$ the stereographic projection of $\mathbf{n}$ from the south pole onto the complex plane. The Majorana polynomial $P_N(z)$ is a monic polynomial of degree $N$, factorizable as $P_N(z) = (-1)^N c_j \prod_{k=1}^N (z - z_k)$ by the fundamental theorem of algebra. The Majorana stars are the points on the unit sphere obtained via the inverse stereographic projection of the roots $z_k$, providing an intuitive depiction of the state's geometric and entanglement properties.
In the context of tri-axis spin squeezed states evolved from the lowest-weight state $|j, -j\rangle$, the representation elucidates how the twisting parameters $\boldsymbol{\mu} = (\mu_0, \mu_1, \mu_2)$ distort the initial degenerate configuration (all stars at the south pole) into asymmetric patterns. For small $j$, explicit forms are
\begin{subequations}
\begin{align}
|1, \boldsymbol{\mu}\rangle &= e^{-i \mu_0 / 2} \left( - \frac{i \xi}{|\xi|} \sin\frac{|\xi|}{2} |1, 1\rangle + \cos\frac{|\xi|}{2} |1, -1\rangle \right), \\
\left| \frac{3}{2}, \boldsymbol{\mu} \right\rangle &= e^{-i 5 \mu_0 / 8} \left( - \frac{\sqrt{3} i \xi}{\vartheta} \sin\frac{\vartheta}{2} \left| \frac{3}{2}, \frac{1}{2} \right\rangle\right. \nonumber\\
&\left.+ \left( \cos\frac{\vartheta}{2} - \frac{i \mu_0}{\vartheta} \sin\frac{\vartheta}{2} \right) \left| \frac{3}{2}, -\frac{3}{2} \right\rangle \right),
\end{align}
\end{subequations}
where $\xi = \mu_1 - i \mu_2$, $|\xi| = \sqrt{\mu_1^2 + \mu_2^2}$, and $\vartheta = \sqrt{\mu_0^2 + 3 |\xi|^2}$. The corresponding Majorana polynomials are
\begin{subequations}
\begin{align}
P_2(z) &= e^{-i \mu_0 / 2} \left( - \frac{i \xi}{|\xi|} \sin\frac{|\xi|}{2} z^2 + \cos\frac{|\xi|}{2} \right), \\
P_3(z) &= e^{-i 5 \mu_0 / 8} \left( - \frac{3 i \xi}{\vartheta} \sin\frac{\vartheta}{2} z^2 + \cos\frac{\vartheta}{2} - \frac{i \mu_0}{\vartheta} \sin\frac{\vartheta}{2} \right).
\end{align}
\end{subequations}
The roots $z_k$ encode the squeezing: for $\boldsymbol{\mu} = 0$, all roots coincide at $z=0$ (south pole); nonzero twisting spreads them, with clustering indicating entanglement.
This representation also facilitates analysis of multipartite entanglement by mapping the symmetric spin-$j$ state to an $N$-qubit symmetric state via $|j, m\rangle \leftrightarrow |S^{(N)}_{j-m}\rangle$, where
\begin{equation}
|S^{(N)}_k\rangle = \frac{1}{\sqrt{\binom{N}{k}}} \sum_{\pi} |\underbrace{0 \cdots 0}_{N-k} \underbrace{1 \cdots 1}_k\rangle_\pi
\end{equation}
sums over all distinct permutations $\pi$. For $j=1$ (two qubits), the tri-axis state is
\begin{equation}
|1, \boldsymbol{\mu}\rangle = e^{-i \mu_0 / 2} \left( - \frac{i \xi}{|\xi|} \sin\frac{|\xi|}{2} |00\rangle + \cos\frac{|\xi|}{2} |11\rangle \right) = \sum_{i,j=0}^1 \Gamma_{ij} |ij\rangle,
\end{equation}
with bipartite concurrence $C = 2 |\det(\Gamma_{ij})| = \sin |\xi|$, quantifying entanglement induced by the twisting~\cite{hill1997bipartite}. For a pure state $|\psi\rangle$, it is defined as the overlap between the state and its "spin-flipped" version: $C(\psi) = |\langle \psi | \tilde{\psi} \rangle|$. For a mixed state $\rho$, the concurrence is calculated using the eigenvalues of a specific matrix related to the spin-flip transformation. The entanglement of formation is then a monotonic function of this concurrence. Similar mappings for higher $j$ reveal how tri-axis interactions generate multipartite entanglement, essential for quantum-enhanced metrology.

In summary, the Majorana stellar representation bridges algebraic and geometric perspectives, enabling deeper insights into the entanglement and squeezing dynamics of tri-axis states, with potential extensions to higher-order correlations and phase transitions.

\subsection{The Classical Phase Space Representation for General Three-Axis Spin Squeezed States}
To gain deeper insight into the dynamics of tri-axis spin squeezed states, we turn to the semiclassical phase space description, valid in the large-$j$ limit where quantum fluctuations become negligible relative to mean-field behavior. This approach maps the collective spin system to a classical vector on the Bloch sphere, revealing fixed points, trajectories, and stability properties that underpin the quantum squeezing mechanisms~\cite{zhang1990coherent, kitagawa1993squeezed}.

In the semiclassical limit, the spin operators are replaced by classical variables $\mathbf{S} = j \mathbf{n}$, where $\mathbf{n} = (\sin\theta \cos\phi, \sin\theta \sin\phi, \cos\theta)$ is a unit vector, and the dynamics follow from the Poisson bracket structure ${S_\alpha, S_\beta} = \epsilon_{\alpha\beta\gamma} S_\gamma$. The effective classical Hamiltonian, obtained by taking expectation values in coherent states and scaling by $1/j$, is
\begin{equation}
\mathcal{H}(\theta, \phi) = \frac{\chi_0}{2} (1 - \cos^2\theta) + \frac{\chi_1}{2} \sin^2\theta \cos 2\phi + \frac{\chi_2}{2} \sin^2\theta \sin 2\phi,
\end{equation}
derived from Eq.~\eqref{SpinHamiltonian} with $\mathbf{J}^2 / j^2 \to 1$. Hamilton's equations govern the motion:
\begin{subequations}
\begin{align}
\dot{\theta} &= \frac{\partial \mathcal{H}}{\partial \phi} = \chi_2 \sin^2\theta \cos 2\phi - \chi_1 \sin^2\theta \sin 2\phi, \\
\dot{\phi} &= -\frac{\partial \mathcal{H}}{\partial \theta} = -(\chi_0 + \chi_1 \cos 2\phi + \chi_2 \sin 2\phi) \sin\theta \cos\theta.
\end{align}
\end{subequations}
These equations describe precessional or nutational orbits on the sphere, analogous to an asymmetric top in rigid-body dynamics.

In the rotated frame of Eq.~\eqref{RotatedHamiltonian}, the phase space simplifies to
\begin{equation}
\mathcal{H}' = \frac{\chi_0}{2} (1 - \cos^2\theta) + \frac{\chi}{2} \sin^2\theta \cos 2\phi,
\end{equation}
highlighting the interplay between uniaxial ($\chi_0$) and biaxial ($\chi$) terms. Fixed points occur where $\dot{\theta} = \dot{\phi} = 0$, corresponding to extrema of $\mathcal{H}'$. For $\chi_0 > \chi$, stable fixed points at the poles ($\theta = 0, \pi$) represent aligned states, with separatrices dividing librational and rotational orbits. As $\chi / \chi_0$ increases, a bifurcation emerges at $\chi = \chi_0$, where equatorial fixed points ($\theta = \pi/2$, $\phi = 0, \pi/2$) become unstable, signaling a classical phase transition analogous to the quantum ESQPT discussed earlier.
Tri-axis squeezed states manifest as Gaussian wavepackets in this phase space, initialized as coherent states centered at $\mathbf{n}_0$ with width $\sim 1/\sqrt{j}$. Under evolution, OAT-dominated regimes ($\chi_0 \gg \chi$) shear the packet along great circles, leading to elliptical spreading and $\xi^2 \sim j^{-2/3}$ scaling. In TAT-dominated cases ($\chi > \chi_0$), twisting induces spiraling distortions, achieving near-Heisenberg-limited squeezing via counter-rotating flows. The full tri-axis dynamics combine these, producing asymmetric deformations tunable by $\chi_2$, enhancing entanglement generation near critical trajectories.

Phase portraits (not shown) for varying $\mu_0 = \chi_0 t$ illustrate this: closed orbits for subcritical parameters evolve into open, chaotic-like paths post-bifurcation, correlating with quantum level clustering in Fig.~\ref{S1}. This classical-quantum correspondence underscores the metrological advantages, as optimal squeezing aligns with minimal phase-space volume along sensitive directions.

In conclusion, the classical phase space elucidates the tunable nature of tri-axis squeezing, bridging mean-field theory with quantum enhancements and paving the way for simulations in large-spin systems.

\subsection{Husimi-$Q$ Quasi-Probability Distribution for One-Axis, Two-Axis, and Three-Axis Spin Squeezed States}
\begin{figure}[tbp]
\begin{center}
\subfloat[$\mu=0.2$\label{sfig:mu02}]{%
\includegraphics[width=0.65\columnwidth]{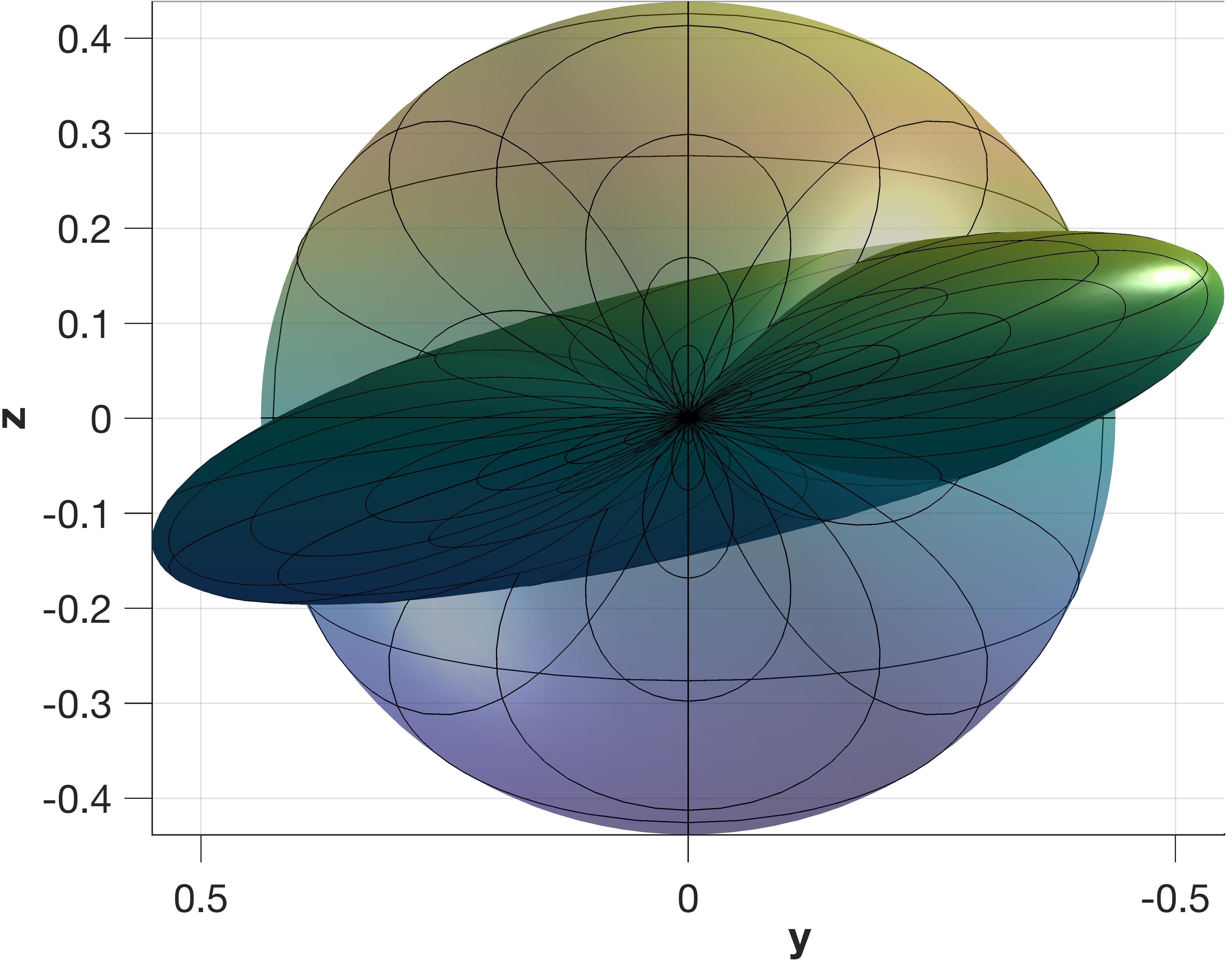}%
}\hfill
\subfloat[$\mu=0.4$\label{sfig:mu04}]{%
\includegraphics[width=0.65\columnwidth]{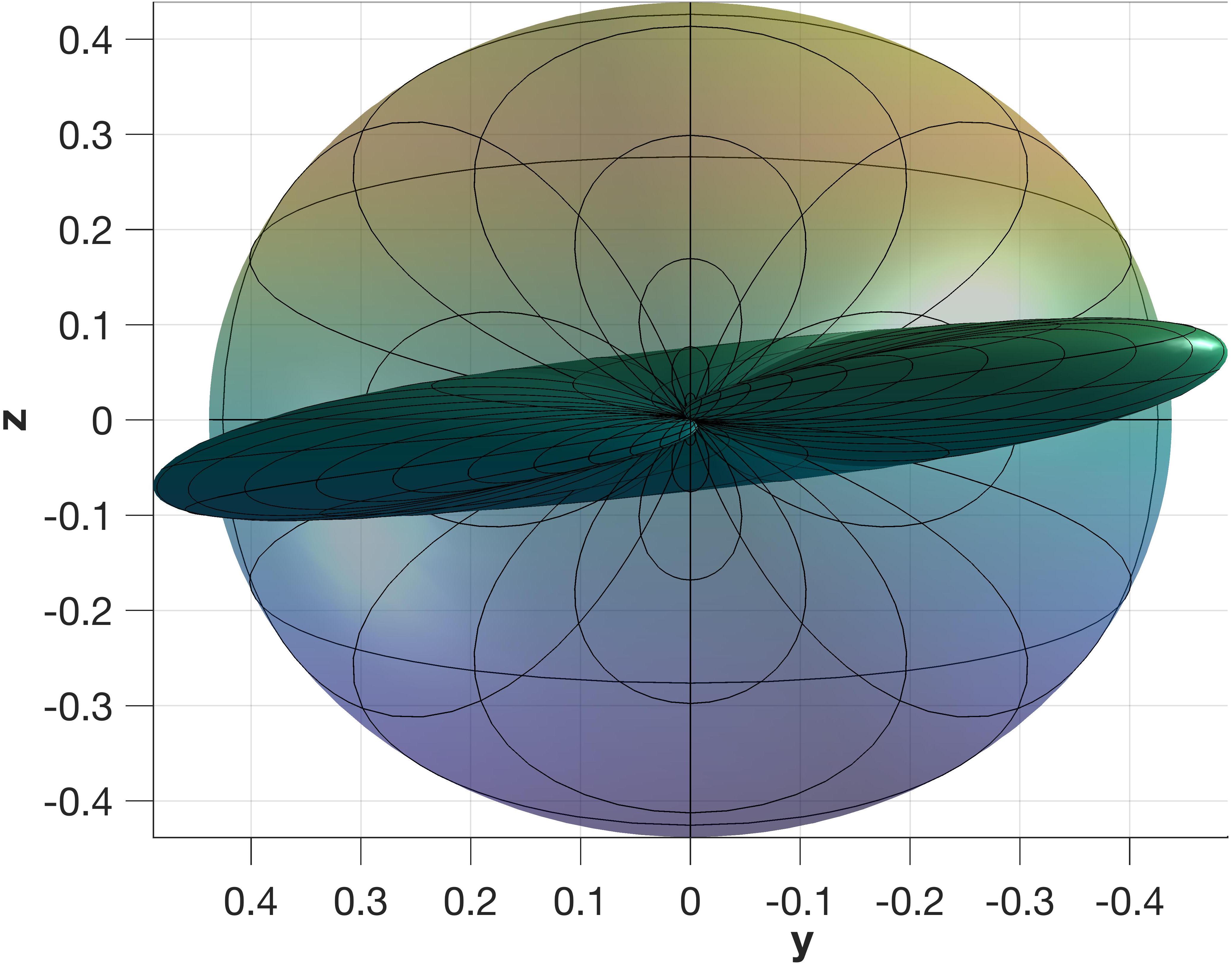}%
}
\caption{Husimi-$Q$ quasi-probability distribution $Q(\theta,\phi)$ for the one-axis spin squeezed state $|j,\mu\rangle$ with $j=20$, where $x\equiv Q(\theta,\phi)\sin\theta\cos\phi$, $y\equiv Q(\theta,\phi)\sin\theta\sin\phi$, and $z\equiv Q(\theta,\phi)\cos\theta$. The spin squeezed states $|j,\mu\rangle$ are shown in viridian, and the initial spin coherent state $|\pi/2,0\rangle$ is depicted in half-transparent moss green and lavender.}
\label{fig:spinsqueezing}
\end{center}
\end{figure}

\begin{figure}[tbp]
\begin{center}
\subfloat[$\nu=0.05$\label{sfig:nu005xy}]{%
\includegraphics[width=0.49\columnwidth]{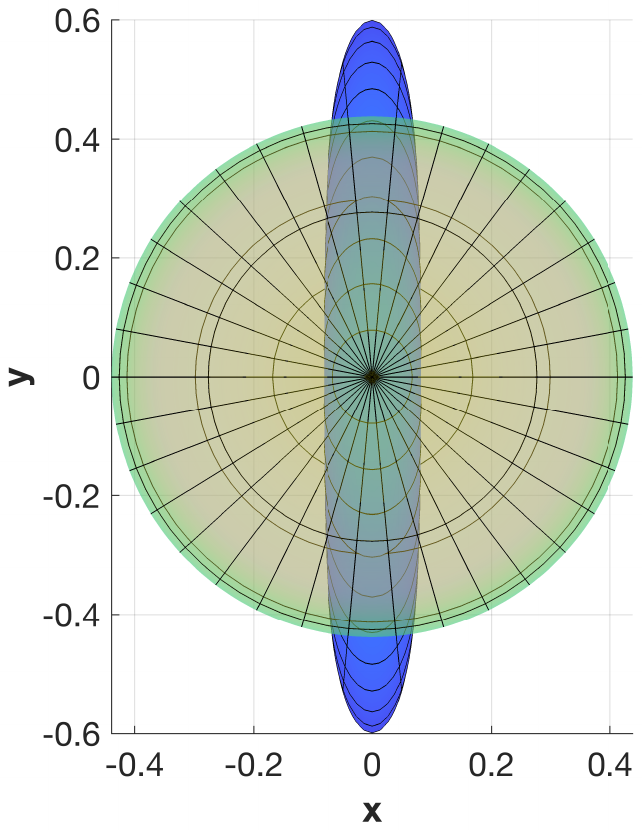}%
}\hfill
\subfloat[$\nu=0.1$\label{sfig:nu01xy}]{%
\includegraphics[width=0.49\columnwidth]{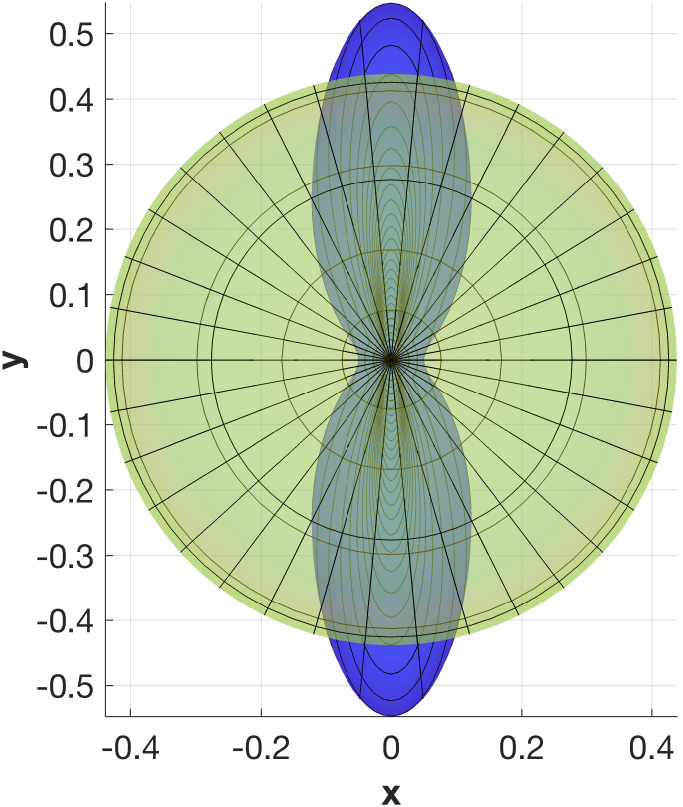}%
}
\hfill
\subfloat[$\nu=0.05$\label{sfig:nu005}]{%
\includegraphics[width=0.50\columnwidth]{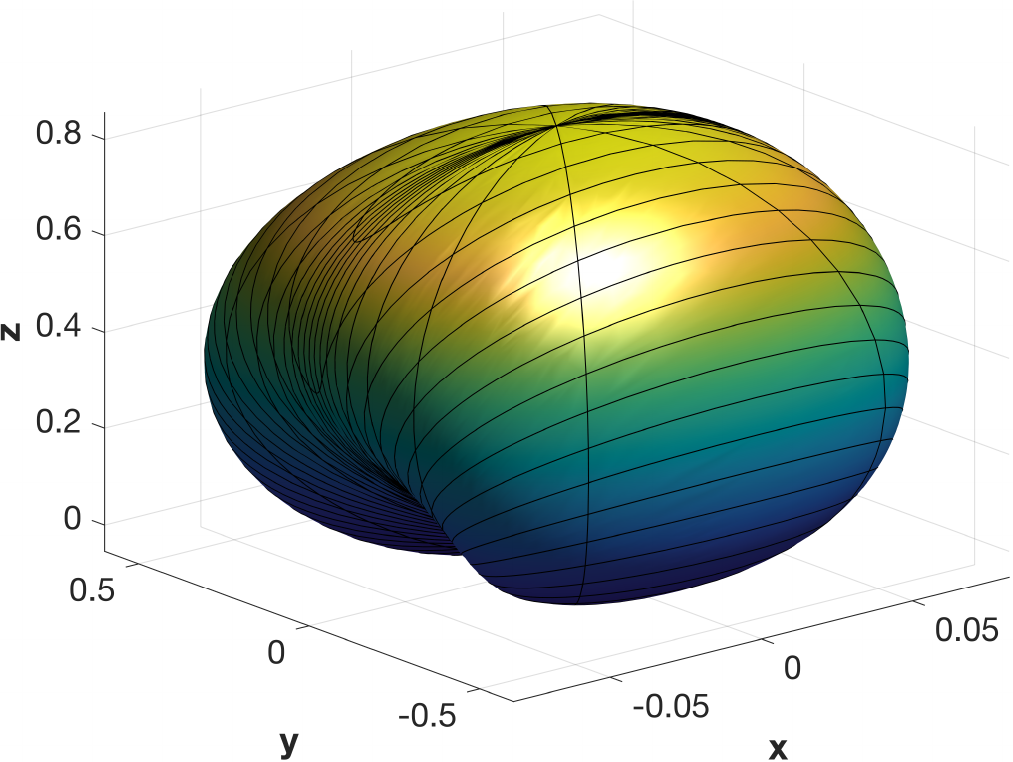}%
}
\hfill
\subfloat[$\nu=0.1$\label{sfig:nu01}]{%
\includegraphics[width=0.50\columnwidth]{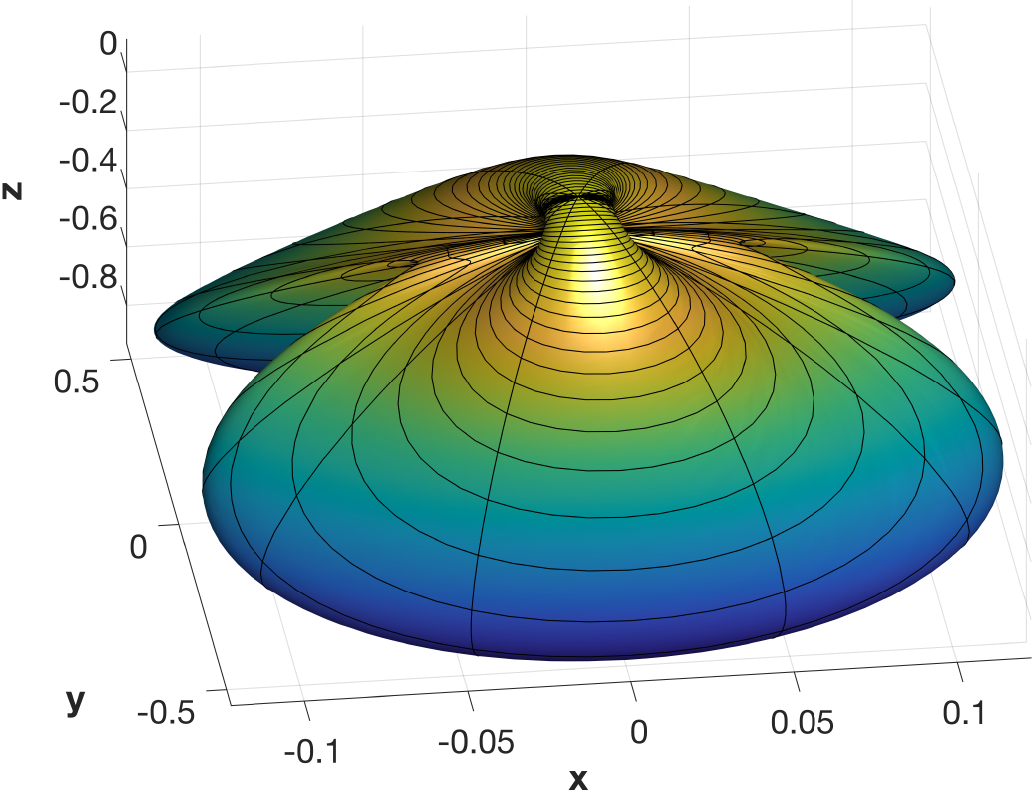}%
}
\caption{Husimi-$Q$ quasi-probability distribution $Q(\theta,\phi)$ for the two-axis spin squeezed state $|j,\nu\rangle$ with $j=20$, where $x\equiv Q(\theta,\phi)\sin\theta\cos\phi$, $y\equiv Q(\theta,\phi)\sin\theta\sin\phi$, and $z\equiv Q(\theta,\phi)\cos\theta$. The upper two panels display $xy$-plane projections of the $Q$ distribution for the two-axis spin squeezed states in blue, with the $Q$ distribution for the initial state $|j,-j\rangle$ shown in half-transparent grass green. The lower two panels present the three-dimensional $Q$ distributions for the two-axis squeezed states.}
\label{fig:twoaxisspinsqueezing}
\end{center}
\end{figure}

The Husimi-$Q$ function offers a crucial method for visualizing quantum states within phase space, defined as the projection onto coherent states to provide a non-negative quasi-probability distribution across the Bloch sphere~\cite{husimi1940some}. Unlike the Wigner function, which may display negativities indicative of nonclassical behavior, the $Q$ function smooths quantum characteristics over a coherent state width, making it an excellent tool for illustrating the anisotropic effects induced by squeezing in collective spin systems~\cite{scully1997quantum}. In this subsection, we systematically explore the $Q$ distributions for one-, two-, and three-axis squeezed states, demonstrating how increasing twisting complexity transforms the initial Gaussian-like profile into elliptical or more sophisticated shapes, reflecting enhanced metrological precision.
We start with the one-axis twisted (OAT) states~\cite{kitagawa1993squeezed},
\begin{equation}
|j, \mu\rangle = e^{-i \mu J_z^2 / 2} \left| \frac{\pi}{2}, 0 \right\rangle = 2^{-j} \sum_{m=-j}^j \sqrt{\binom{2j}{j+m}} e^{-i \mu m^2 / 2} |j, m\rangle,
\end{equation}
initialized along the $x$-axis. The corresponding Husimi-$Q$ function is
\begin{align}
Q(\theta, \phi) &= \frac{2j+1}{4\pi} \left| \langle \theta, \phi | j, \mu \rangle \right|^2 \nonumber \\
&= \frac{2j+1}{4\pi \cdot 2^{2j}} \left| \sum_{m=-j}^j \binom{2j}{j+m} \left( \cos\frac{\theta}{2} \right)^{j-m} \left( e^{i\phi} \sin\frac{\theta}{2} \right)^{j+m} e^{-i \mu m^2 / 2} \right|^2.
\end{align}
Figure~\ref{fig:spinsqueezing} presents $Q(\theta, \phi)$ for $j=20$ at $\mu = 0.2$ and $0.4$, depicted in Cartesian coordinates where $x = Q \sin\theta \cos\phi$, $y = Q \sin\theta \sin\phi$, and $z = Q \cos\theta$. When $\mu=0$, the distribution exhibits spherical symmetry, characteristic of a coherent state. As $\mu$ becomes nonzero, shearing occurs, reshaping the profile into an ellipsoid with disrupted rotational symmetry. The degree of squeezing strengthens with increasing $\mu$, and the optimal squeezing direction shifts over time, a defining trait of OAT dynamics that restricts scalability to $\xi^2 \sim N^{-2/3}$.
Next, we investigate two-axis counter-twisted (TACT) states, which achieve superior Heisenberg-limited scaling $\xi^2 \sim N^{-1}$ through balanced orthogonal twistings~\cite{kitagawa1993squeezed}. These states are generated as
\begin{equation}
|j, \nu\rangle = e^{-\nu (J_+^2 - J_-^2)/2} |j, -j\rangle = \sum_{m=-j}^j c_m |j, m\rangle,
\end{equation}
where $c_{-j+n} = 0$ for odd $n$ due to parity conservation. The $Q$ function is
\begin{equation}
Q(\theta, \phi) = \frac{2j+1}{4\pi} \left| \sum_{m=-j}^j \sqrt{\binom{2j}{j+m}} \left( \cos\frac{\theta}{2} \right)^{j-m} \left( \sin\frac{\theta}{2} e^{i\phi} \right)^{j+m} c_m \right|^2.
\end{equation}
Fig.~\ref{fig:twoaxisspinsqueezing} displays $Q(\theta, \phi)$ for $j=20$ at $\nu = 0.05$ and $0.1$. The upper panels present $xy$-plane projections, showing consistent squeezing along the $y$-axis and antisqueezing along the $x$-axis, invariant with $\nu$—a significant improvement over OAT. The lower panels illustrate the full 3D distributions, evolving from a pea-like shape to a clam-like configuration as $\nu$ increases from 0.05 to 0.1, indicating uniform, directionally focused squeezing that intensifies along $y$ while broadening in perpendicular directions.
Finally, we extend our analysis to the tri-axis squeezed states, which integrate OAT and TACT within the anisotropic LMG framework. These states evolve under the full Hamiltonian in Eq.~\eqref{SpinHamiltonian} as
\begin{equation}
|j, \boldsymbol{\mu}\rangle = \exp\left\{ - \frac{i}{2} \left[ \mu_0 (\mathbf{J}^2 - J_z^2) + \mu_1 (J_x^2 - J_y^2) + \mu_2 (J_x J_y + J_y J_x) \right] \right\} |j, -j\rangle,
\end{equation}
with $\boldsymbol{\mu} = (\mu_0, \mu_1, \mu_2)$. The corresponding $Q$ function is
\begin{equation}
Q(\theta, \phi) = \frac{2j+1}{4\pi} \left| \langle \theta, \phi | j, \boldsymbol{\mu} \rangle \right|^2,
\end{equation}
computed through expansion in Dicke states or numerical techniques for finite $j$. In the tri-axis regime, the distribution displays asymmetric ellipticities adjustable by the ratios $\mu_1/\mu_0$ and $\mu_2/\mu_0$, blending OAT shearing with TACT spiraling. For instance, a dominant $\mu_0$ produces rotating ellipsoids, while balanced $\mu_1$ and $\mu_2$ yield twisted, clam-like shapes with heightened concurrence near critical points. This flexibility enhances metrological performance, approaching $\sin|\xi| \approx 1$ for low $j$, and supports quantum rotor simulations.

In conclusion, the Husimi-$Q$ distributions progressively unveil how multi-axis twisting enriches the squeezing landscape, transitioning from simple ellipsoids to complex asymmetries, thus highlighting the tri-axis model's potential for advanced quantum applications.

\subsection{Majorana Stellar Representation for the One-Axis, Two-Axis, and Three-Axis Spin Squeezed States}
The Majorana stellar representation offers a geometric visualization of pure spin-$j$ states by mapping them to constellations of $2j$ points on the Bloch sphere, revealing underlying symmetries, entanglement patterns, and phase-space distributions~\cite{majorana1932atomi, bloch1946nuclear, wick2004majorana}. This approach is particularly insightful for spin squeezed states, where the deformation of the stellar configuration from an initial clustered arrangement illustrates the effects of nonlinear twisting, providing a complementary perspective to quasiprobability functions and aiding in the analysis of metrological enhancements and multipartite correlations~\cite{chabaud2017stellar}.

For one-axis spin squeezed states $|j, \mu\rangle$, the associated stellar polynomial can be explicitly formulated for arbitrary $N=2j$, where $q \equiv e^{i \mu / 2}$ is a unit complex number with the squeezing parameter encoded in its phase (see App.~B). The normalized Majorana polynomial is
\begin{equation}
F_N(z) \equiv 2^j q^{j^2} P_N(z) = \sum_{n=0}^N \binom{N}{n} q^{-n^2 + n N} (-z)^n,
\end{equation}
such that $F_N(1) = 1$ for any natural number $N$. Direct computation yields the first ten polynomials for $1 \leq N \leq 10$: $F_1(z) = 1 - z$, $F_2(z) = 1 - 2 q z + z^2$, and
\begin{align}
F_3(z) &= 1 - 3 q^2 z + 3 q^2 z^2 - z^3, \nonumber \\
F_4(z) &= 1 - 4 q^3 z + 6 q^4 z^2 - 4 q^3 z^3 + z^4, \nonumber \\
F_5(z) &= 1 - 5 q^4 z + 10 q^6 z^2 - 10 q^6 z^3 + 5 q^4 z^4 - z^5, \nonumber \\
F_6(z) &= 1 - 6 q^5 z + 15 q^8 z^2 - 20 q^9 z^3 + 15 q^8 z^4 - 6 q^5 z^5 + z^6, \nonumber \\
F_7(z) &= 1 - 7 q^6 z + 21 q^{10} z^2 - 35 q^{12} z^3 + 35 q^{12} z^4 \nonumber \\
&\quad - 21 q^{10} z^5 + 7 q^6 z^6 - z^7, \nonumber \\
F_8(z) &= 1 - 8 q^7 z + 28 q^{12} z^2 - 56 q^{15} z^3 + 70 q^{16} z^4 - 56 q^{15} z^5 \nonumber \\
&\quad + 28 q^{12} z^6 - 8 q^7 z^7 + z^8, \nonumber \\
F_9(z) &= 1 - 9 q^8 z + 36 q^{14} z^2 - 84 q^{18} z^3 + 126 q^{20} z^4 \nonumber \\
&\quad - 126 q^{20} z^5 + 84 q^{18} z^6 - 36 q^{14} z^7 + 9 q^8 z^8 - z^9, \nonumber \\
F_{10}(z) &= 1 - 10 q^9 z + 45 q^{16} z^2 - 120 q^{21} z^3 \nonumber \\
&\quad + 210 q^{24} z^4 - 252 q^{25} z^5 + 210 q^{24} z^6 \nonumber \\
&\quad - 120 q^{21} z^7 + 45 q^{16} z^8 - 10 q^9 z^9 + z^{10}.
\end{align}
When $\mu = 0$ or $q = 1$, these reduce to $(1 - z)^N$, with all roots degenerate at $z = 1$, corresponding to stars clustered at the antipodal point. Nonzero $\mu$ spreads the roots, forming arcs or loops on the sphere that rotate with time, reflecting the shearing dynamics of OAT and the associated $\xi^2 \sim N^{-2/3}$ scaling.

Extending to two-axis counter-twisted states $|j, \nu\rangle = e^{-\nu (J_+^2 - J_-^2)/2} |j, -j\rangle$, the Majorana polynomial captures the biaxial deformation. Due to parity conservation, the roots exhibit even symmetry, often pairing antipodally or forming symmetric constellations around the equator. For small $\nu$, the stars bifurcate from the south pole into two clusters twisting oppositely, evolving into ring-like distributions at optimal squeezing times, which visualize the Heisenberg-limited noise reduction and enhanced entanglement compared to OAT~\cite{kitagawa1993squeezed}.

For the general three-axis squeezed states $|j, \boldsymbol{\mu}\rangle$, the stellar representation unifies these behaviors, with the polynomial roots depending on the interplay of $\mu_0$, $\mu_1$, and $\mu_2$. In the rotated frame, the configuration interpolates between uniaxial arcs and biaxial rings, displaying asymmetric clustering tunable by anisotropy. Near critical ratios (e.g., $\chi_0 \approx \chi$), the stars exhibit chaotic spreading, correlating with ESQPTs and boosted concurrence up to $\sin|\xi| \approx 1$, as seen in low-$j$ explicit forms.

In essence, the Majorana stellar framework provides a unified geometric lens for multi-axis squeezing, illuminating transitions from coherent to highly entangled regimes and guiding optimizations for quantum sensing.

Experimental realizations of spin squeezing have been achieved across various quantum platforms, including atomic ensembles, trapped ions, cavity quantum electrodynamics (QED) systems, and Rydberg atom arrays, leveraging nonlinear interactions to surpass the standard quantum limit in precision measurements~\cite{ma2011quantum, hosten2016measurement}. Among these, Bose-Einstein condensates (BECs) in double-well potentials stand out for their tunability and ability to exhibit rich many-body phenomena, such as self-trapping transitions and associated topological features. These systems provide a controllable environment to probe the interplay between quantum geometry, nonlinear dynamics, and entanglement, with direct implications for quantum metrology and simulation.

A paradigmatic example is the observation of macroscopic quantum self-trapping (MQST) in BECs confined to double-well traps, first demonstrated by Albiez et al. in 2005 using $^{87}$Rb atoms~\cite{albiez2005direct}. In this setup, the condensate is split into two weakly coupled modes by a tunable optical barrier, forming an asymmetric double-well potential. The system is described by the two-mode Bose-Hubbard Hamiltonian,
\begin{equation}
H = -J (a_L^\dagger a_R + a_R^\dagger a_L) + \frac{U}{2} (n_L^2 + n_R^2) + \Delta n_L,
\end{equation}
where $a_{L/R}^\dagger$ ($a_{L/R}$) create (annihilate) atoms in the left/right well, $n_{L/R} = a_{L/R}^\dagger a_{L/R}$ are number operators, $J$ is the tunneling strength, $U$ is the on-site interaction energy, and $\Delta$ accounts for any asymmetry. For weak interactions ($U N / J \ll 1$, where $N$ is the total atom number), the dynamics exhibit Rabi-like oscillations between wells. However, as $U N / J$ exceeds a critical value (typically around 2 for symmetric wells), the system transitions to MQST, where the population imbalance self-localizes due to nonlinear repulsion, suppressing tunneling.

This self-trapping transition corresponds to an excited-state quantum phase transition (ESQPT) in the mean-field limit, characterized by a bifurcation in phase space from oscillatory to fixed-point dynamics~\cite{chuchem2010quantum}. The critical point manifests as level clustering in the energy spectrum, as illustrated in Fig.~\ref{S1} for a related tri-axis model, where eigenvalues $E_k$ versus normalized interaction $\mu_0$ show avoided crossings and density singularities near $\mu_0^c \approx 1.5$. In BEC experiments, similar spectral features have been inferred from dynamical responses, with ESQPTs driving enhanced fluctuations and entanglement generation~\cite{relano2008excited}.

Complementing these dynamics, spin squeezing emerges in multicomponent or spinor BECs, where internal degrees of freedom enable effective twisting Hamiltonians. For instance, in two-component $^{87}$Rb BECs (e.g., hyperfine states $|F=1, m_F=-1\rangle$ and $|F=2, m_F=1\rangle$), collisional interactions generate one-axis twisting, as realized by Gross et al. in 2010, achieving up to 8 dB of squeezing~\cite{gross2010nonlinear}. Extending to double-well geometries, Faraday-imaging techniques induce measurement-based squeezing, as demonstrated by Li et al. in 2021, where nondestructive probes in a $^{87}$Rb condensate yielded spin squeezing while preserving coherence~\cite{li2021faraday}.

A deeper connection arises in the work of Kam and Liu (2017), who explored ESQPTs in two-component BECs with self-trapping, revealing topological signatures in the quantum geometry~\cite{kam20172+}. Here, the Berry curvature—a gauge field quantifying the geometric phase accumulated during adiabatic parameter variation—diverges at the bifurcation surface, forming a conic singularity analogous to a magnetic monopole in momentum space~\cite{berry1984quantal}. Mathematically, for a parameter-dependent Hamiltonian $H(\mathbf{R})$, the Berry curvature for state $|n(\mathbf{R})\rangle$ is
\begin{equation}
\Omega_{ij}^{(n)} = i \left( \langle \partial_{R_i} n | \partial_{R_j} n \rangle - \langle \partial_{R_j} n | \partial_{R_i} n \rangle \right),
\end{equation}
with singularities at critical points where eigenstates become degenerate or the gauge structure breaks down. In BEC self-trapping, this leads to non-integrable phase accumulation, manifesting as anomalous transport or Aharonov-Bohm-like effects in cyclic evolutions~\cite{williamson2017experimental}.

Experimental probes of such geometric phases in ultracold gases have advanced significantly, with direct measurements of Berry curvature via anomalous Hall drift in optical lattices, as reported by Aidelsburger et al. in 2015 and later refined in 2017~\cite{aidelsburger2015measuring, asteria2019measuring}. In double-well BECs, similar techniques could map the curvature divergence, linking ESQPTs to topological metrology. Recent extensions to spin-orbit-coupled BECs further enhance squeezing via synthetic gauge fields, achieving up to 10 dB reduction in spin noise~\cite{luo2016tunable}.

These realizations underscore the versatility of BECs for simulating quantum many-body physics, with self-trapping and squeezing enabling applications in atomic clocks, gravimeters, and quantum simulators~\cite{peik1997bloch}. Future directions include integrating cavity QED for collective enhancement or Rydberg interactions for long-range twisting, potentially realizing three-axis squeezing in hybrid platforms.

\section{Conclusion}
In this work, we have introduced and comprehensively analyzed three-axis spin squeezed states within the framework of the anisotropic Lipkin-Meshkov-Glick (LMG) model, extending the foundational one- and two-axis twisting paradigms proposed by Kitagawa and Ueda~\cite{kitagawa1993squeezed}. By incorporating direction-dependent quadratic couplings and external fields, our model unifies uniaxial and biaxial regimes into a tunable asymmetric quantum rotor, enabling elliptical quasiprobability distributions and enhanced multipartite entanglement. Through derivations of the Hamiltonian, semiclassical Euler-top dynamics, and phase-space representations—including Majorana stellar constellations and Husimi-$Q$ functions—we have demonstrated optimal squeezing scalings of $\xi^2 \sim N^{-2/3}$ for one-axis twisting and Heisenberg-limited $\xi^2 \sim 1/N$ for two-axis variants, with three-axis states achieving boosted metrological gains and concurrence approaching $\sin|\xi| \approx 1$ in low-$j$ systems.

A central finding is the emergence of second-order ground-state quantum phase transitions (QPTs) between paramagnetic and ferromagnetic phases, driven by anisotropy tuning, with critical exponents $\gamma=1$ and adherence to Kibble-Zurek universality in quench dynamics at $\lambda_z = \lambda_{x,y}$. Furthermore, the energy spectrum reveals excited-state QPTs (ESQPTs) marked by level clustering and singularities in the density of states, as evidenced in Fig.~\ref{S1}, where eigenvalues $E_k$ exhibit avoided crossings near the critical parameter $\mu_0^c \approx 1.5$. These critical phenomena, analyzed via classical phase portraits and quantum geometric measures like Berry curvature divergences, underscore the model's rich interplay between integrability, chaos, and topological features.

Our results build upon the established connections between spin squeezing and many-body criticality in the LMG model~\cite{lipkin1965validity, meshkov1965validity, glick1965validity}, advancing the field by incorporating three-axis interactions that allow for finer control over asymmetry and entanglement. The semiclassical analogies to rigid-body rotations provide intuitive insights into the squeezing mechanisms, where OAT induces uniaxial shearing, TACT yields biaxial distortions, and tri-axis dynamics facilitate hybrid behaviors with tunable ellipticities. This generalization addresses limitations in prior models, such as the time-dependent optimal squeezing angle in OAT, by enabling invariant directions in mixed regimes, as visualized in the Husimi-$Q$ distributions transitioning from ellipsoids to clam-like shapes.

However, challenges remain, including the exact solvability for arbitrary $j$ in the tri-axis case, which may require advanced techniques like the algebraic Bethe ansatz~\cite{pan2017exact, pan2017exact2}. Finite-size effects and decoherence in realistic implementations could also temper the predicted scalings, necessitating further numerical simulations using tensor network methods or exact diagonalization for larger $N$. Comparisons with related systems, such as the Dicke model for superradiance~\cite{wang1973phase}, highlight shared ESQPT signatures but emphasize the LMG's advantage in all-to-all connectivity for scalable squeezing.

The implications of this work are multifaceted, spanning quantum metrology, many-body theory, and topological quantum physics. In metrology, the enhanced squeezing and concurrence in tri-axis states promise improved sensitivity in atomic clocks, magnetometers, and interferometers, potentially surpassing current benchmarks in platforms like Rydberg arrays and cavity-QED~\cite{ma2011quantum}. The association with ESQPTs suggests novel protocols for critical sensing, where operating near phase boundaries amplifies parameter estimation via divergent susceptibilities~\cite{santos2016excited}.

Theoretically, our findings deepen the understanding of classical-quantum correspondences in nonlinear systems, with the conic singularities in Berry curvature mirroring gauge field monopoles in BEC self-trapping transitions~\cite{kam20172+}. This bridges quantum optics with condensed matter physics, offering a platform to simulate topological effects in non-Hermitian or driven systems. Practically, the model's integrability in limiting cases facilitates exact benchmarks for quantum simulators, aiding the design of fault-tolerant quantum devices.

Looking ahead, several avenues warrant exploration. Experimentally, realizing tri-axis twisting in ultracold atoms via spin-dependent lattices or microwave dressing could validate the predicted ESQPTs and squeezing enhancements, building on recent advances in Faraday-imaging and spin-orbit coupling~\cite{li2021faraday, luo2016tunable}. Theoretically, extending the model to include disorder, finite-range interactions, or higher-order nonlinearities may reveal quantum chaos indicators, such as out-of-time-order correlators, and their impact on squeezing resilience.

Integrating machine learning for optimizing twisting parameters or classifying phase transitions could accelerate applications. Additionally, investigating geometric phases in adiabatic cycles near ESQPTs may uncover novel topological invariants, with potential links to quantum computing via entangled state preparation. Ultimately, this work paves the way for harnessing multi-axis squeezing in next-generation quantum technologies, fostering interdisciplinary progress in precision measurement and many-body simulation.

\begin{appendix}

\section{Spin Squeezing Parameter for Spin States}
For spin coherent states, the variance $(\Delta\hat{J}_{\boldsymbol{n}})^2$ of the spin operator $\hat{J}_{\boldsymbol{n}}\equiv \hat{J}\cdot\boldsymbol{n}$ varies for different directions $\boldsymbol{n}$. However, it remains the same along the directions $\boldsymbol{n}_{\perp}$ perpendicular to the mean spin direction, i.e., $(\Delta\hat{J}_{\boldsymbol{n}_{\perp}})^2=j/2$ \cite{perelomov2012generalized, kam2023coherent}. Kitagawa and Ueda argued that, the criteria for spin squeezing is that the variance $(\Delta\hat{J}_{\boldsymbol{n}_{\perp}})^2<j/2$ for some specific directions $\boldsymbol{n}_{\perp}$ perpendicular to the mean spin direction \cite{kitagawa1993squeezed}. In other words, one may define the spin squeezing parameter $\xi_s$ via
\begin{equation}
\xi_s^2\equiv\frac{\mbox{min}((\Delta\hat{J}_{\boldsymbol{n}_{\perp}})^2)}{j/2},
\end{equation}
where the minimization takes place over all directions $\boldsymbol{n}_{\perp}$ perpendicular to the mean spin direction. Let us denote the mean spin direction for a general spin-$j$ state as $\boldsymbol{n}_0\equiv(\theta,\phi)$. For example, the mean spin direction for the one-axis spin squeezed state $|j,\mu\rangle\equiv e^{-\frac{i}{2}\mu\hat{J}_z^2}|\pi/2,0\rangle$ is the $x$-direction. With respect to the mean spin direction $\boldsymbol{n}_0$, the other two orthonormal bases are
\begin{subequations}
\begin{align}
\boldsymbol{n}_1&\equiv(-\sin\phi,\cos\phi,0),\\
\boldsymbol{n}_2&\equiv(\cos\theta\cos\phi,\cos\theta\sin\phi,-\sin\theta),
\end{align}
\end{subequations}
where the above expressions are only valid for $\theta\neq 0$ or $\pi$. For $\theta=0$ or $\pi$, the mean spin direction is along the $\pm z$ directions, and $\phi$ can be fixed as $0$ or $\pi$ respectively. An arbitrary direction perpendicular to the mean spin direction can be represented as $\boldsymbol{n}_{\perp}\equiv \boldsymbol{n}_1\cos\varphi+\boldsymbol{n}_2\sin\varphi$. As $\langle\hat{J}_{\boldsymbol{n}_1}\rangle=\langle\hat{J}_{\boldsymbol{n}_2}\rangle=0$, the variance $\langle\hat{J}_{\boldsymbol{n}_\perp}^2\rangle$ can be written as
\begin{align}
\langle\hat{J}_{\boldsymbol{n}_\perp}^2\rangle&=\langle\hat{J}_{\boldsymbol{n}_1}^2\rangle\cos^2\varphi+\langle\{\hat{J}_{\boldsymbol{n}_1},\hat{J}_{\boldsymbol{n}_2}\}\rangle\cos\varphi\sin\varphi+\langle\hat{J}_{\boldsymbol{n}_2}^2\rangle\sin^2\varphi\nonumber\\
&=\frac{1}{2}\left(\langle\hat{J}_{\boldsymbol{n}_1}^2+\hat{J}_{\boldsymbol{n}_2}^2\rangle\right)+\frac{A}{2}\cos 2\varphi+\frac{B}{2}\sin 2\varphi,
\end{align}
where $A\equiv\langle\hat{J}_{\boldsymbol{n}_1}^2-\hat{J}_{\boldsymbol{n}_2}^2\rangle$, $B\equiv 2\Cov(\hat{J}_{\boldsymbol{n}_1},\hat{J}_{\boldsymbol{n}_2})\equiv \langle\{\hat{J}_{\boldsymbol{n}_1},\hat{J}_{\boldsymbol{n}_2}\}\rangle$ and $\Cov(\hat{J}_{\boldsymbol{n}_1},\hat{J}_{\boldsymbol{n}_2})$ is the covariance between $\hat{J}_{\boldsymbol{n}_1}$ and $\hat{J}_{\boldsymbol{n}_2}$. Hence, one immediately obtains
\begin{equation}
\frac{(\Delta\hat{J}_{\boldsymbol{n}_{\perp}})^2}{j/2}\geq\frac{1}{j}\left(\langle\hat{J}_{\boldsymbol{n}_1}^2+\hat{J}_{\boldsymbol{n}_2}^2\rangle-\sqrt{A^2+B^2}\right),
\end{equation}
where the variance $(\Delta\hat{J}_{\mathbf{n}_{\perp}})^2$ reaches its minimum when the relations $\cos2\varphi=-A/\sqrt{A^2+B^2}$ and $\sin 2\varphi\equiv -B/\sqrt{A^2+B^2}$ are satisfied. In other words, the spin squeezing parameter is determined by
\begin{equation}
\xi_s^2=\frac{1}{j}\left\{\langle\hat{J}_{\boldsymbol{n}_1}^2+\hat{J}_{\boldsymbol{n}_2}^2\rangle-\sqrt{\left(\langle\hat{J}_{\boldsymbol{n}_1}^2-\hat{J}_{\boldsymbol{n}_2}^2\rangle\right)^2+4\Cov(\hat{J}_{\boldsymbol{n}_1},\hat{J}_{\boldsymbol{n}_2})^2}\right\},
\end{equation}
where the optimal squeezing angle is given by
\begin{equation}
\varphi=\left\{\begin{aligned}
 	 \frac{1}{2}\arccos\left(\frac{-A}{\sqrt{A^2+B^2}}\right)\:\:\mbox{for}\:\:B\leq 0,\\
  	\pi- \frac{1}{2}\arccos\left(\frac{-A}{\sqrt{A^2+B^2}}\right)\:\:\mbox{for}\:\:B>0.
	\end{aligned}
	\right.
\end{equation}

\section{One-axis and two-axis spin squeezed states}
\subsection{One-axis spin squeezed states}
In line with Kitagawa and Ueda \cite{kitagawa1993squeezed}, one-axis spin squeezed states $|j,\zeta,\mu\rangle\equiv e^{-i\mu J_z^2/2}|j,\zeta\rangle$ are generated from the quadratic spin Hamiltonian $H\equiv \chi J_z^2/2$, where $\mu\equiv \chi t$ is a squeezing parameter, and $|j,\zeta\rangle\equiv e^{\zeta J_+-\zeta^* J_-}|j,-j\rangle$ is an initial spin coherent state \cite{perelomov2012generalized, kam2023coherent}. In the Heisenberg picture, the spin operators $J_x$, $J_y$ and $J_z$ obey the precession equations
\begin{equation}\label{OneAxisHeisenberg}
\dot{J}_x=\frac{i\chi}{2}(J_zJ_y+J_yJ_z),\dot{J}_y=-\frac{i\chi}{2}(J_zJ_x+J_xJ_z),\dot{J}_z=0.
\end{equation}
From Eq.\:\eqref{OneAxisHeisenberg}, one sees that the precession of the spin is modulated by $J_z$, which is a twisting effect analogous to the self-phase modulation in nonlinear optical waveguides \cite{agrawal2019nonlinear}. As the nonlinear twisting Hamiltonian $H=\chi J_z^2/2$ is along the $z$-axis, one may choose the initial state as a spin coherent state along the $x$-axis, so that the one-axis spin squeezed state becomes
\begin{equation}
|j,\mu\rangle\equiv e^{-\frac{i}{2}\mu J_z^2}\left|\frac{\pi}{2},0\right\rangle= 2^{-j}\sum_{m=-j}^{j}\sqrt{C_{2j}^{j+m}}e^{-\frac{i}{2}\mu m^2} |j,m\rangle.
\end{equation}
A direct calculation shows that the mean spin direction of the one-axis spin squeezed state $|j,\mu\rangle$ is along the $x$-axis, i.e., $\langle J_x\rangle = j\cos^{2j-1}\frac{\mu}{2}$, $\langle J_y\rangle=\langle J_z\rangle=0$, but the magnitude of the mean spin vector is reduced by a factor $\cos^{2j-1}\frac{\mu}{2}$, compared to that of the initial spin coherent state $|\pi/2,0\rangle$. A direct computation yields the variances and covariances of the spin operators with respect to the one-axis spin squeezed states
\begin{align}\label{VarianceOneAxis}
&(\Delta J_x)^2=\frac{j}{2}\left[j+\frac{1}{2}+(j-\frac{1}{2})\cos^{2j-2}\mu-2j\cos^{2(2j-1)}\frac{\mu}{2}\right],\nonumber\\
&(\Delta J_y)^2=\frac{j}{2}\left[j+\frac{1}{2}-(j-\frac{1}{2})\cos^{2j-2}\mu\right],(\Delta J_z)^2=\frac{j}{2},\nonumber\\
&\Cov(J_y,J_z)=j(j-\frac{1}{2})\sin\frac{\mu}{2}\cos^{2j-2}\frac{\mu}{2},\nonumber\\
&\Cov(J_z,J_x)=\Cov(J_x,J_y)=0,
\end{align}
Notice that, in the absence of the nonlinear twisting effects, i.e., $\mu=0$, one immediately recovers the variances for the spin coherent states, $(\Delta J_x)^2=0$, $(\Delta J_y)^2=(\Delta J_z)^2=j/2$. As the mean spin direction of the one-axis spin squeezed states $|j,\mu\rangle$ is along the $x$-axis, one may evaluate the variance of $J_{\boldsymbol{n}_{\perp}}\equiv J_z\cos\theta+J_y\sin\theta$, where $\boldsymbol{n}_{\perp}$ is perpendicular to the mean spin direction. As $\langle J_z\rangle=\langle J_y\rangle=0$, the variance of $J_{\boldsymbol{n}_{\perp}}$ becomes
\begin{equation}
(\Delta J_{\boldsymbol{n}_{\perp}})^2 = \frac{A}{2}\cos 2\theta + \frac{B}{2}\sin 2\theta +\frac{C}{2},
\end{equation}
where $C=j/2+(\Delta J_y)^2$, $A=j/2-(\Delta J_y)^2$, $B\equiv 2\Cov(J_y,J_z)$. In particular, for a spin coherent state with $\mu=0$, one obtains $A=B=0$ and $C=j$. Hence, the variance $(\Delta J_{\boldsymbol{n}_{\perp}})^2$ for a spin coherent state is always equal to $j/2$. In general, one obtains
\begin{equation}
(\Delta J_{\boldsymbol{n}_{\perp}})^2 =(\Delta J_{\boldsymbol{n}_{\perp}})^2_{\textrm{max}}\cos^2\theta_\varphi+(\Delta J_{\boldsymbol{n}_{\perp}})^2_{\textrm{min}}\sin^2\theta_\varphi,
\end{equation}
where $\theta_\varphi\equiv \theta-\varphi/2$, $\tan\varphi\equiv B/A$, and $(\Delta J_{\boldsymbol{n}_{\perp}})^2_{\textrm{max}}$ and $(\Delta J_{\boldsymbol{n}_{\perp}})^2_{\textrm{min}}$ are the maximal and minimal variances given by
\begin{subequations}
\begin{align}
(\Delta J_{\boldsymbol{n}_{\perp}})^2_{\textrm{max}}&=\frac{1}{2}(C+\sqrt{A^2+B^2}),\\
(\Delta J_{\boldsymbol{n}_{\perp}})^2_{\textrm{min}}&=\frac{1}{2}(C-\sqrt{A^2+B^2}).
\end{align}
\end{subequations}
The minimum value is reached when $\theta-\varphi/2=\pi/2$, which comprises the phase-matching condition for one-axis spin squeezed states. From Eq.\:\eqref{VarianceOneAxis}, one immediately obtains the spin squeezing parameter for the one-axis spin squeezed states
\begin{align}
\xi_s^2&=1+\frac{N-1}{2}\left(\frac{1-\cos^{N-2}\mu}{2}\right.\nonumber\\
&\left.-\sqrt{\frac{(1-\cos^{N-2}\mu)^2}{4}+4\sin^2\frac{\mu}{2}\cos^{2(N-2)}\frac{\mu}{2}}\right),
\end{align}
where $N\equiv 2j$. Clearly, for $j=1/2$ or $N=1$, the spin squeezing parameter always equals to $1$, and hence spin squeezing effects only exist in higher spin systems with $j>1/2$. Here, the spin squeezing parameters $\xi_s$ for one-axis spin squeezed states $|j,\mu\rangle$ with $j\leq 2$ are shown as explicit examples:
\begin{subequations}
\begin{align}
\xi_s^2&=1-\left|\sin\frac{\mu}{2}\right|\geq 0,\:\mbox{for}\:j=1,\\
\xi_s^2&=1+\sin^2\frac{\mu}{2}-\sqrt{\sin^4\frac{\mu}{2}+\sin^2\mu}\geq \frac{1}{3},\:\mbox{for}\:j=\frac{3}{2},\\
\xi_s^2&=1+3\cos^2\frac{\mu}{2}\left(\sin^2\frac{\mu}{2}-\sqrt{\sin^4\frac{\mu}{2}+\sin^2\frac{\mu}{2}}\right)\\\nonumber
&\geq 1+3(1-r)(r-\sqrt{r^2+r})\approx 0.3025,\:\mbox{for}\:j=2,
\end{align}
\end{subequations}
where $r\approx 0.2228$ is the smallest positive root of the cubic equation $8r^3+5r^2-6r+1=0$. 

For one-axis spin squeezed states $|j,\mu\rangle$, the associated Majorana polynomials are given by
\begin{equation}
P_{2j}(z)=2^{-j}\sum_{m=-j}^j C_{2j}^{j+m}q^{-m^2}(-z)^{j+m}.
\end{equation}
where $q\equiv e^{\frac{i}{2}\mu}$ is a unit complex number with the squeezing parameter being its phase. To simplify the discussion, one can introduce the normalized Majorana polynomial
\begin{equation}
F_N(z)\equiv 2^jq^{j^2}P_N(z) = \sum_{n=0}^N C_N^n q^{-n^2+nN}(-z)^n,
\end{equation}
such that $F_N(1)=1$ for any $N\in \mathbb{N}+$. A direct computation yields the first few normalized Majorana polynomials for $1\leq N\leq 10$: $F_1(z) = 1-z$, $F_2(z) = 1 - 2qz + z^2$ and
\begin{align}
F_3(z) &= 1 - 3q^2z + 3q^2z^2-z^3,\nonumber\\
F_4(z) &= 1 - 4q^3z+ 6q^4z^2 - 4q^3z^3 + z^4,\nonumber\\
F_5(z) &= 1 - 5q^4z+ 10q^6z^2 -10q^6z^3+ 5q^4z^4 - z^5,\nonumber\\
F_6(z) &= 1 - 6q^5z+ 15q^8 z^2- 20q^9 z^3+ 15q^8z^4- 6q^5z^5+ z^6,\nonumber\\
F_7(z) &= 1 - 7q^6z+ 21q^{10} z^2- 35q^{12} z^3+35q^{12}z^4\nonumber\\
&- 21q^{10}z^5+ 7q^6z^6-z^7,\nonumber\\
F_8(z) &= 1 - 8q^7z + 28q^{12}z^2 - 56q^{15}z^3+ 70q^{16}z^4- 56q^{15}z^5 \nonumber\\
&+ 28q^{12}z^6  - 8q^7z^7+ z^8,\nonumber\\
F_9(z) &= 1 - 9q^8z + 36q^{14}z^2 - 84q^{18}z^3+ 126q^{20}z^4 \nonumber\\
&- 126q^{20}z^5+ 84q^{18}z^6  - 36q^{14}z^7+ 9q^8z^8-z^9,\nonumber\\
F_{10}(z) &= 1 - 10q^9z + 45q^{16}z^2 - 120q^{21}z^3 \nonumber\\
&+ 210q^{24}z^4 - 252q^{25}z^5+210q^{24}z^6 \nonumber\\
& - 120q^{21}z^7 + 45q^{16}z^{8} -10q^{9}z^{9} +z^{10}.
\end{align}
Clearly, for $\mu=0$ or $q=1$, the Majorana polynomials become $(1-z)^N$ for arbitrary $N$, which contains $N$ identical roots at $z=1$.

\subsection{Two-axis spin squeezed states}
Unlike the one-axis spin squeezed states $|j,\mu\rangle$, there is another type of spin squeezed states, namely the two-axis spin squeezed states, of which the optimal squeezing angles are invariant under time evolution \cite{ma2011quantum}. The two-axis spin squeezed states can be generated by applying nonlinear twisting simultaneously clockwise and counterclockwise along the two orthonormal axes in the $\theta=\pi/2$ and $\phi=\pm\pi/4$ directions, which are both in the plane normal to the mean spin direction. Hence, the nonlinear twisting Hamiltonians for the two-axis spin squeezed states $|j,\nu\rangle\equiv e^{-it\hat{H}(\hat{J}_+,\hat{J}_-)}|j,-j\rangle=e^{-\frac{\nu}{2}(\hat{J}_+^2-\hat{J}_-^2)}|j,-j\rangle$ are given by $\hat{H}\equiv \chi(\hat{J}_x\hat{J}_y+\hat{J}_y\hat{J}_x)=\frac{\chi}{2i}(\hat{J}_+^2-\hat{J}_-^2)$, where $\nu\equiv\chi t$. Without loss of generality, one can assume that $\chi$ is positive, and thus $\nu$ is also positive.

One characteristic of the two-axis spin squeezed states is that the co-variance $\langle\hat{J}_x\hat{J}_y+\hat{J}_x\hat{J}_y\rangle$ vanishes, as $\hat{J}_x\hat{J}_y+\hat{J}_x\hat{J}_y$ is conserved in time, i.e., $[\hat{J}_x\hat{J}_y+\hat{J}_x\hat{J}_y, \hat{H}]=0$, and the co-variances $\langle\hat{J}_\pm^2\rangle$ vanish for the initial lowest weight state $|j,-j\rangle$. Moreover, as $\hat{J}_\pm^2|j,m\rangle\propto |j,m\pm 2\rangle$, the two-axis twisting Hamiltonian change $m$ in the Dicke states $|j,m\rangle$ by $2$. Hence, the two-axis spin squeezed state $|j,\nu\rangle$ must be an even parity state, i.e., $|j,\nu\rangle$ is spanned only by Dicke states $|j,-j+n\rangle$ with $n$ even, and the states $\hat{J}_\pm|j,\nu\rangle$ must be odd parity states, i.e., $\hat{J}_{\pm}|j,\nu\rangle$ are spanned only be Dicke states $|j,-j+n\rangle$ with $n$ odd. In other words, the two-axis spin squeezed states $|j,\nu\rangle$ as well as $\hat{J}_\pm^2|j,\nu\rangle$ are eigenstates of the parity operator $(-1)^{\hat{J}_z+j}$ with an eigenvalue $1$, while the states $\hat{J}_\pm|j,\nu\rangle$ are the eigenstate of the parity operator $(-1)^{\hat{J}_z+j}$ with eigenvalues $-1$. In this regard, both the expectations $\langle\hat{J}_x\rangle\equiv\frac{1}{2}(\langle \hat{J}_+\rangle+\langle \hat{J}_-\rangle)$ and $\langle\hat{J}_y\rangle\equiv \frac{1}{2i}(\langle\hat{J}_+\rangle -\langle\hat{J}_-\rangle)$ vanish, as they involve inner products between Dicke states with different parities. Unlike the one-axis spin squeezed state, as both the expectations $\langle\hat{J}_x\rangle$ and $\langle\hat{J}_y\rangle$ vanish, the mean spin direction always points along the $z$-direction. For the same reason, both the expectations $\langle\hat{J}_\pm\hat{J}_z\rangle$ and $\langle\hat{J}_z\hat{J}_\pm\rangle$, as well as the co-variances $\langle\hat{J}_x\hat{J}_z+\hat{J}_z\hat{J}_x\rangle$ and $\langle\hat{J}_y\hat{J}_z+\hat{J}_z\hat{J}_y\rangle$ vanish identically. 

As a remark, the two-axis twisting model was thought to be analytically unsolvable for arbitrary values of $j$. In fact, it can be solved for both integer and half-integer $j$ by using the algebraic Bethe ansatz, which is based on the Jordan-Schwinger map and the Fock-Bargmann representation \cite{pan2017exact, pan2017exact2}. But this generally involved complicated series expressions. In the next section, we present explicit expressions for the two-axis spin squeezed states for $j\leq 5/2$, along with the associated variances.

For $j=1$, the two-axis spin squeezed state is a superposition of the lowest weight state and the highest weight state
\begin{equation}
    |1,\nu\rangle = -\sin\nu|1,1\rangle+\cos\nu|1,-1\rangle.
\end{equation}
A direct computation yields the expectations $\langle J_x\rangle = \langle J_y\rangle = 0$ and $\langle J_z \rangle = -\cos 2\nu$, along with the following expectations
\begin{equation}
    \langle J_x^2\rangle=\frac{1}{2}(1-\sin 2\nu),\langle J_y^2\rangle=\frac{1}{2}(1+\sin 2\nu),\langle J_z^2\rangle=1.
\end{equation}
From this one immediately obtains the following variances
\begin{equation}
    (\Delta J_z)^2=\sin^2 2\nu,
    (\Delta J_{\boldsymbol{n}_{\perp}})^2= \frac{1}{2}(1-\sin 2\nu\cos 2\varphi),
\end{equation}
where $J_{\boldsymbol{n}_{\perp}}\equiv J_x\cos\varphi+J_y\sin\varphi$. It immediately yields the spin squeezing parameter
\begin{equation}
    \xi_s^2=1-|\sin 2\nu|,
\end{equation}
Clearly, the minimum value of $(\Delta J_{\boldsymbol{n}_{\perp}})^2$ occurs at $\varphi =0$ for $k\pi<\nu<(k+1/2)\pi$, and at $\varphi=\pi/2$ for $(k+1/2)\pi<\nu<(k+1)\pi$, where $k$ is an integer. In particular, $\Delta J_x$ vanishes when $\nu=(1/4+k)\pi$, $\Delta J_y$ vanishes when $\nu=(3/4+k)\pi$, and $\Delta J_z$ vanishes when $\nu=k\pi/2$. In this case, the Majorana polynomial is given by
\begin{equation}
P_2(z)=-\sin\nu z^2+\cos\nu,
\end{equation}
where the two Majorana zeros are located at $z_1=\sqrt{\cot\nu}$ and $z_2=-\sqrt{\cot\nu}$. It is evident that they converse only when $\nu=(2k+1)\pi/2$, where $k$ is an integer.

For $j=3/2$, the two-axis spin squeezed state is a superposition of states that share the same parity as the lowest weight state, along with the lowest weight state itself
\begin{equation}
    |\frac{3}{2},\nu\rangle=-\sin u|\frac{3}{2},\frac{1}{2}\rangle+\cos u|\frac{3}{2},-\frac{3}{2}\rangle.
\end{equation}
where $u\equiv \sqrt{3}\nu$. A direct computation yields the expectations    
\begin{subequations}
\begin{gather}
    \langle J_x\rangle = \langle J_y\rangle = 0,\langle J_z\rangle = \frac{1}{2}-2\cos^2 u,\\
    \langle J_x^2\rangle=\frac{5}{4}-\cos \theta_-,\langle J_y^2\rangle=\frac{5}{4}-\cos \theta_+,\\
    \langle J_z^2\rangle=\frac{5}{4}+\cos 2u,
\end{gather}
\end{subequations}
where $\theta_\pm\equiv 2u\pm \pi/3$, which satisfy $\cos\theta_++\cos\theta_-=\cos 2u$. From this one immediately obtains the variances    
\begin{gather}   
    (\Delta J_z)^2=\sin^2 2u,\\
    (\Delta J_{\boldsymbol{n}_{\perp}})^2= \frac{5}{4}-\frac{1}{2}\cos 2u-\frac{\sqrt{3}}{2}\sin 2u\cos 2\varphi,
\end{gather}
where $J_{\boldsymbol{n}_{\perp}}\equiv J_x\cos\varphi+J_y\sin\varphi$. It immediately yields the spin squeezing parameter
\begin{equation}
    \xi_s^2=\frac{5}{3}-\frac{4}{3}\cos(2u\mp\frac{\pi}{3})\geq \frac{1}{3},
\end{equation}
where the minus sign is for $k\pi<u<(k+1/2)\pi$, and the plus sign is for $(k+1/2)\pi<u<(k+1)\pi$. Clearly, the minimum value of $(\Delta J_{\boldsymbol{n}_{\perp}})^2$ occurs at $\varphi =0$ for $k\pi<u<(k+1/2)\pi$, and at $\varphi=\pi/2$ for $(k+1/2)\pi<u<(k+1)\pi$. In particular, $\Delta J_z$ vanishes when $u=k\pi/2$. In this case, the Majorana polynomial is given by
\begin{equation}
P_3(z)=\cos u -\sqrt{3}\sin u z^2,
\end{equation}
where the two Majorana zeros are located at 
\begin{equation}
z_1=\sqrt{\frac{\cot u}{\sqrt{3}}},z_2=-\sqrt{\frac{\cot u}{\sqrt{3}}},z_3=\infty.
\end{equation} 

For $j=2$, the two-axis spin squeezed state is again a superposition of states that have the same parity as the lowest weight state, along with the lowest weight state itself
\begin{equation}
|2,\nu\rangle=\sin^2u|2,2\rangle-\frac{\sin 2u}{\sqrt{2}}|2,0\rangle+\cos^2u|2,-2\rangle,
\end{equation}
where $u\equiv \sqrt{3}\nu$. A direct computation yields the expectations $\langle J_x\rangle = \langle J_y\rangle =0$ and $\langle J_z\rangle = -2\cos 2u$, along with the following expectations   
\begin{subequations}
\begin{align}
    \langle J_x^2\rangle&=1+\sin^22u-\sqrt{3}\sin2u,\\
    \langle J_y^2\rangle&=1+\sin^22u+\sqrt{3}\sin2u,\\
    \langle J_z^2\rangle&=3+\cos 4u.
\end{align}
\end{subequations}
From this one immediately obtains the variances    
\begin{subequations}
\begin{align}
    (\Delta J_z)^2&= 1-\cos 4u,\\
    (\Delta J_{\boldsymbol{n}_{\perp}})^2&=1+\sin^2u-\sqrt{3}\sin 2u\cos 2\varphi.
\end{align}
\end{subequations}
where $J_{\boldsymbol{n}_{\perp}}\equiv J_x\cos\varphi+J_y\sin\varphi$. It immediately yields the spin squeezing parameter
\begin{equation}
\xi_s^2=\frac{1}{4}+\left(\sin 2u\mp\frac{\sqrt{3}}{2}\right)^2\geq \frac{1}{4}.
\end{equation}
where the minus sign is for $k\pi<u<(k+1/2)\pi$, and the plus sign is for $(k+1/2)\pi<u<(k+1)\pi$. Clearly, the minimum value of $(\Delta J_{\boldsymbol{n}_{\perp}})^2$ occurs at $\varphi =0$ for $k\pi<u<(k+1/2)\pi$, and at $\varphi=\pi/2$ for $(k+1/2)\pi<u<(k+1)\pi$. In particular, $\Delta J_z$ vanishes when $u=k\pi/2$. In this case, the Majorana polynomial is given by
\begin{equation}
P_4(z)=\sin^2 u z^4-\sqrt{3}\sin 2u z^2+\cos^2u,
\end{equation}
where the four Majorana zeros are located at 
\begin{align}
z_k=\pm\sqrt{\sqrt{3}\pm 2\sqrt{2}}\sqrt{\cot u},k=1,2,3,4.
\end{align} 
Clearly, the four Majorana zeros are non-degenerate in general, and they become degenerate only when $\cot u=0$, i.e., $u=\pi/2+n\pi$.

For $j=5/2$, the two-axis spin squeezed state is again a superposition of states that share the same parity as the lowest weight state, along with the lowest weight state itself
\begin{equation}
    |\frac{5}{2},\nu\rangle=a|\frac{5}{2},\frac{3}{2}\rangle+b|\frac{5}{2},-\frac{1}{2}\rangle+c|\frac{5}{2},-\frac{5}{2}\rangle,
\end{equation}
where
\begin{equation*}
a\equiv \frac{3\sqrt{5}}{7}\sin^2u, b\equiv -\sqrt{\frac{10}{7}}\sin u\cos u, c\equiv 1-\frac{5}{7}\sin^2u
\end{equation*}
and $u\equiv \sqrt{7}\nu$. A direct computation yields the expectations $\langle J_x\rangle = \langle J_y\rangle =0$ and $\langle J_z\rangle = \frac{1}{2}(3a^2-b^2-5c^2)$, along with the following expectations
\begin{subequations}
\begin{align}
\langle J_x^2 \rangle&= \frac{1}{4}\left(3a^2+12b^2+5\right)+3\sqrt{2}ab+\sqrt{10}bc,\\
\langle J_y^2 \rangle&= \frac{1}{4}\left(3a^2+12b^2+5\right)-3\sqrt{2}ab-\sqrt{10}bc,\\
\langle J_z^2 \rangle&= \frac{1}{4}\left(9a^2+b^2+25c^2\right).
\end{align}
\end{subequations}
From this one immediately obtains the variance
\begin{equation}
    (\Delta J_{\boldsymbol{n}_{\perp}})^2=\frac{3a^2}{4}+3b^2+\frac{5}{4}+\sqrt{2}b(3a+\sqrt{5}c)\cos 2\varphi,
\end{equation}  
where $J_{\boldsymbol{n}_{\perp}}\equiv J_x\cos\varphi+J_y\sin\varphi$. A direct computation yields
\begin{align}
\frac{2}{j}(\Delta J_{\boldsymbol{n}_{\perp}})^2 & = \frac{27}{49}\sin^4u +\frac{6}{7}\sin^22u+1\nonumber\\
&-\frac{4}{7}\sin 2u \left(1+\frac{4}{7}\sin^2u\right)\cos 2\varphi.
\end{align}
In this case, the Majorana polynomial is given by
\begin{equation}
P_5(z)=\frac{15}{7}\sin^2 u z^4-\frac{5}{\sqrt{7}}\sin 2u z^2+1-\frac{5}{7}\sin^2 u.
\end{equation}
The five Majorana zeros are located at 
\begin{equation}
z_k = \pm \sqrt{\frac{\sqrt{7}}{3}\left(\cot u \pm\sqrt{\cot^2 u -\frac{9}{4}\csc^2 u +\frac{3}{7}}\right)},
\end{equation}
where $k=1,2,3,4$ and $z_5=\infty$.

\section{Explicit expressions for the three-axis spin squeezed states, the variances and the covariances}
For an initial lowest-weight state $|j,-j\rangle$, the three-axis spin squeezed states are given by
\begin{equation}
|j,\mu_0,\mu\rangle\equiv \exp\left\{-\frac{i}{2}[\mu_0(\boldsymbol{J}^2-J_z^2)+\mu(J_x^2-J_y^2)]\right\}|j,-j\rangle.
\end{equation}
As $J_x^2-J_y^2=(J_+^2+J_-^2)/2$ and $J_{\pm}^2|j,m\rangle\propto |j,m\pm 2\rangle$, the three-axis spin squeezed state $|j,\mu_0,\mu\rangle$ is spanned only by Dicke states $|j,-j+n\rangle$ with $n$ even. Hence, one immediately obtains $\langle J_x\rangle = \langle J_y\rangle = 0$ and $\Cov(J_y,J_z)=\Cov(J_z,J_x)=0$. In the following, we evaluate the explicit expressions for the three-axis spin squeezed states $|j,\mu_0,\mu\rangle$ for the first few $j$, as well as the variances and the covariances of the spin operators. 

For $j=1$, a direct computation yields
\begin{subequations}
\begin{gather}
|1,\mu_0,\mu\rangle = e^{-\frac{1}{2}\mu_0 i}(-i\sin\frac{\mu}{2}|1,1\rangle + \cos\frac{\mu}{2}|1,-1\rangle),\\
\langle J_z\rangle =1, (\Delta J_z)^2 = 0,(\Delta J_x)^2 = (\Delta J_y)^2 = \frac{1}{2},\\
\Cov(J_x,J_y)=\frac{\sin\mu}{2},\xi_s^2 = \frac{1}{2}(1-|\sin\mu|).
\end{gather}
\end{subequations}

For $j=3/2$, a direct computation yields
\begin{subequations}
\begin{gather}
|\frac{3}{2},\mu_0,\mu\rangle =c_{\frac{1}{2}}|\frac{3}{2},\frac{1}{2}\rangle + c_{-\frac{1}{2}}|\frac{3}{2},-\frac{1}{2}\rangle,\\
c_{\frac{1}{2}}  =  -e^{-\frac{5}{4}\mu_0 i}\frac{\sqrt{3}i\mu}{\theta}\sin\frac{\theta}{2},\\
c_{-\frac{1}{2}}= e^{-\frac{5}{4}\mu_0 i}(\cos\frac{\theta}{2}+\frac{i\mu_0}{\theta}\sin\frac{\theta}{2}),
\end{gather}
\end{subequations}
where $\theta\equiv \sqrt{\mu_0^2+3\mu^2}$. Hence, the variances and covariances of $J_x$, $J_y$ and $J_z$ with respect to the three-axis spin squeezed states are
\begin{subequations}
\begin{gather}
\langle J_z\rangle = \frac{3}{2}(4\chi^2-1),(\Delta J_z)^2 =12\chi^2(1-3\chi^2),\\
(\Delta J_x)^2 =\frac{3}{4}+\frac{3\mu(\mu-\mu_0)}{\theta^2}\sin^2\frac{\theta}{2},\\
(\Delta J_y)^2 =\frac{3}{4}+\frac{3\mu(\mu+\mu_0)}{\theta^2}\sin^2\frac{\theta}{2},\\
\Cov(J_x,J_y)=\frac{3\mu}{\theta}\sin\frac{\theta}{2}\cos\frac{\theta}{2},\\
\xi_s ^2 = 1+2\chi^2-4\chi\sqrt{1-3\chi^2},\chi\equiv \frac{|\mu|}{\theta}\sin\frac{\theta}{2},
\end{gather}
\end{subequations}
where the spin squeezing parameter takes the minimum value $(4-\sqrt{13})/3\approx 0.131$ at
\begin{equation}
\chi = \sqrt{\frac{1}{6}-\frac{1}{6\sqrt{13}}} \approx 0.347.
\end{equation}

\end{appendix}

\end{document}